\documentclass[aps,pra,preprint]{revtex4-1}
\usepackage[colorlinks=true, linkcolor=blue, citecolor=blue, urlcolor=blue]{hyperref}
\usepackage{natbib}
\usepackage{graphicx} 
\usepackage{subfigure}
\usepackage{tabularx,stackengine}
\usepackage{bm}
\usepackage{amssymb,amsmath}
\usepackage{tikz}
\usetikzlibrary {decorations.pathmorphing, decorations.pathreplacing, decorations.shapes,arrows.meta}

\begin{document}

\title{Non-Hermitian Band Topology and Edge States \\ in Atomic Lattices}
\author{Wenxuan Xie}
\affiliation{Department of Applied Physics, Yale University, New Haven, CT 06511, USA}
\email{wenxuan.xie@yale.edu}
\author{John C. Schotland}
\affiliation{Department of Mathematics and Department of Physics, Yale University, New Haven, CT 06511, USA} 
\email{john.schotland@yale.edu}

\begin{abstract}
We investigate the band structure and topological phases of one- and two-dimensional bipartite atomic lattices mediated by long-range dissipative radiative coupling. By deriving an effective non-Hermitian Hamiltonian for the single-excitation sector, we demonstrate that the low-energy dynamics of the system are governed by a Dirac equation with a complex Fermi velocity. We analyze the associated topological invariants for both the SSH and honeycomb models, utilizing synthetic gauge fields to break time-reversal symmetry in the latter. Finally, we explicitly verify the non-Hermitian bulk-edge correspondence by deriving analytical solutions for edge states localized at domain boundaries.
\end{abstract}

\maketitle 

\section{\label{sec:introdution}INTRODUCTION}
Many-body problems in quantum optics provide a rich setting in which to model the phenomena of condensed matter physics. The ability to realize optical analogs of exotic quantum phases, especially those arising in topological materials, is of considerable recent interest~\cite{aidelsburger2013quantum, rechtsman2013photonic, kim2017topological, mancini2015observation, stefanatos2018engineering, leonard2023realization, jotzu2014experimental}. Such phases can be created in atomic lattices, in which quantum emitters interact with quantized electromagnetic fields. Research in this direction has spanned a variety of lattice types and geometries. In one-dimensional (1D) systems, the Su-Schrieffer-Heeger (SSH) model, characterized by staggered hopping amplitudes within a diatomic unit cel, serves as the simplest setting in which topological edge states arise \cite{su1979solitons, cooper2019topological, jenkins2016observation}. Two-dimensional (2D) systems, such as the honeycomb lattice, give rise to even richer phenomena, including Dirac cones, chiral edge states, and robust topological excitations \cite{jacqmin2014direct, wu2015topological, percik2023non, xue2020non, goldman2016topological, savoia2017many, jenkins2012controlled, perczel2017photonicA, perczel2017topological, perczel2020optically}.

In this paper, we investigate the interaction between a quantized optical field and lattices composed of two distinct species of two-level atoms. In such systems, interatomic coupling is mediated by a 3D free-space field, resulting in interactions that are both long-ranged and radiative. The long-ranged nature of the interactions poses a significant challenge for determining the band structure, as it requires evaluating slowly convergent lattice sums \cite{antezza2009fano, antezza2009spectrum, shahmoon2017cooperative, perczel2017photonicA, perczel2017topological, perczel2020optically}. While a closed-form expression exists for the 1D lattice sum, such expressions are generally elusive for 2D systems. Previous computational efforts for 2D lattices have relied on regularization techniques \cite{perczel2017photonicA, perczel2017topological, perczel2020optically} or Fourier integral representations of the Green's function \cite{antezza2009fano, antezza2009spectrum, shahmoon2017cooperative}. However, these approaches are numerically inefficient and necessitate the introduction of an artificial cutoff. To address these limitations, we utilize the theta function transform and Ewald summation \cite{borwein2013lattice}. This technique maps the 2D lattice sum onto a 1D integral of Jacobi theta functions and a series of exponentially convergent sums, providing a numerically efficient solution without requiring an artificial cutoff.

We begin by deriving an effective two-band Hamiltonian for the atomic excitations. Due to the radiative nature of the photon-mediated coupling, the effective Hamiltonian is intrinsically non-Hermitian. Using the derived expressions for the lattice sums, we compute the band structures for both 1D and 2D lattices, demonstrating the emergence of Dirac points. Beyond numerical analysis, we provide a  proof based on lattice symmetry arguments that establishes the universality of these Dirac points. We demonstrate that the physics near resonance is governed by a Dirac equation characterized by a uniform Fermi velocity. This finding is robust: it holds regardless of whether the interactions are short- or long-ranged and Hermitian or non-Hermitian, provided that the underlying lattice symmetry is preserved. As a consequence, our results are broadly applicable to diverse experimental setups, including waveguide and circuit QED, even when photons are spatially confined. In the specific context of the lattice models we study, the Fermi velocity is complex-valued, allowing for the study of non-Hermitian phenomena, such as the non-Hermitian skin effect \cite{yao2018edge, kunst2018biorthogonal, lee2019anatomy, okuma2020topological, gong2018topological, lin2023topological, tanaka2024non, hu2024topological, amelio2024mathematical, jana2023emerging}.

Finally, we explore the topological properties of the 1D and 2D systems. The non-Hermiticity of the effective Hamiltonian requires the use of non-Hermitian topological band theory \cite{ashida2020non, gong2018topological, li2022gain, shen2018topological, ding2022non, ammari2024mathematical, davies2025two}. For the 1D SSH model, a non-trivial topological phase emerges in the presence of chiral symmetry. In the 2D honeycomb lattice, a non-trivial topology requires breaking time-reversal symmetry. Given that the atoms are charge-neutral and do not couple directly to a gauge field, we implement synthetic gauge fields, inspired by existing proposals for atomic systems \cite{dalibard2011colloquium, kennedy2015observation, bardyn2016topological, roushan2017chiral}, to break time-reversal symmetry, resulting in non-trivial topological phases. We conclude by verifying the bulk-edge correspondence through for a domain-wall problem associated with the Dirac equation, establishing a direct link between the bulk topological invariant and the emergence of protected edge states.

The paper is structured as follows. Sec. \ref{sec:model} introduces the model under investigation and presents the derivation of the effective two-band Hamiltonian. Next, Sec. \ref{sec:SSH} analyzes the 1D SSH model, focusing on its band structure, winding number, and edge states. Sec. \ref{sec:honeycomb} examines the band structure and topological characteristics of the 2D honeycomb lattice. A summary and concluding remarks are provided in Sec. \ref{sec:discuss}. The appendices contain derivations of the lattice sums and the Dirac equation.


\section{\label{sec:model}MODEL}
\subsection{\label{subsec:hamiltonian}Model Hamiltonian}

We consider a system of two-level atoms interacting with the quantized electromagnetic field in three-dimensional space. While we employ a scalar field model for simplicity, our methods readily generalize to the full vector electromagnetic field.
The atoms are arranged in either a one-dimensional chain or a two-dimensional honeycomb lattice. The lattices are composed of  two species of two-level atoms, labeled $A$ and $B$, which are assumed to be sufficiently far apart so that short-range interatomic interactions
can be neglected. The Hamiltonian of the system is of the form
\begin{align} 
   \hat H =  &\sum\limits_{\mathbf{k}} \hbar \omega_{\mathbf{k}} \hat{a}_{\mathbf{k}}^{\dagger} \hat{a}_{\mathbf{k}}
    + \sum\limits_{\alpha =  A, B}\sum\limits_{j}  \hbar \Omega_{\alpha} \hat{\sigma}_{\alpha j}^{\dagger} \hat{\sigma}_{\alpha j} 
    + \sum\limits_{\alpha =  A, B} \sum\limits_{ j, \mathbf{k}} \hbar g_{\alpha} \left( {e}^{\mathrm{i} \mathbf{k} \cdot \mathbf{r}_{\alpha j}} \hat{\sigma}_{\alpha j}^{\dagger} \hat{a}_{\mathbf{k}} 
    + {e}^{-\mathrm{i}\mathbf{k} \cdot \mathbf{r}_{\alpha j}} \hat{\sigma}_{\alpha j} \hat{a}^{\dagger}_{\mathbf{k}}\right) ,
    \label{eq:hamiltonian} 
\end{align}
where we have imposed the rotating wave and dipole approximations. 
The three terms in Eq.~\eqref{eq:hamiltonian} correspond to the Hamiltonians of the electromagnetic field, the atoms, and their interaction, respectively. In addition, the operator
$\hat{a}_{\mathbf{k}}^{\dagger}$ ($\hat{a}_{\mathbf{k}}$) creates (annihilates) a photon with frequency $\omega_{\mathbf{k}}=c|\mathbf k|$ and wavevector $\mathbf{k}$. We denote by $\hat{\sigma}^{\dagger}_{\alpha j}$ ($\hat{\sigma}_{\alpha j}$) the atomic raising (lowering) operator for an atom of type $\alpha \in \{A, B\}$ at the point $\mathbf{r}_{\alpha j}$ in the $j$th unit cell, with resonance frequency $\Omega_{\alpha}$. The coupling strength for each atomic species is given by $g_{\alpha}$ and is assumed to frequency independent.

We restrict our attention to the single-excitation subspace, where the total number of excitations—comprising either one excited atom or one photon—is conserved, consistent with the rotating wave approximation. A stationary state in this subspace is defined by
\begin{equation} 
\lvert \psi \rangle = \left( \sum\limits_{j} \psi_{Aj} \hat{\sigma}_{Aj}^{\dagger} + \sum\limits_{j} \psi_{Bj} \hat{\sigma_{Bj}}^{\dagger} + \sum\limits_{\mathbf{k}} c_{\mathbf{k}} \hat{a}_{\mathbf{k}}^{\dagger} \right) \lvert 0 \rangle, 
\label{eq:state} 
\end{equation}
where $\lvert 0 \rangle$ denotes the combined ground state of the atoms and the field. The coefficients $\psi_{\alpha j}$ and $c_{\mathbf{k}}$ are the probability amplitudes of  exciting an atom of type $\alpha$ in cell $j$ and creating a photon with wavevector $\mathbf{k}$.

The state $\lvert \psi \rangle$ obeys the time-independent Schrödinger equation $\hat H \lvert \psi \rangle = \hbar \omega \lvert \psi \rangle$. It follows that  $\psi_{\alpha j}$ and $c_{\mathbf{k}}$ obey the algebraic equations
\begin{subequations} 
    \begin{align} 
    &\omega \psi_{Aj} = \Omega_{A}\psi_{Aj}  + g_{A} \sum_{\mathbf{k}} {e}^{\mathrm{i} \mathbf{k} \cdot \mathbf{r}_{A j} } c_{\mathbf{k}},  \\ 
    &\omega \psi_{Bj}  = \Omega_{B}\psi_{Bj} + g_{B} \sum_{\mathbf{k}} {e}^{\mathrm{i} \mathbf{k} \cdot \mathbf{r}_{Bj} } c_{\mathbf{k}},\\ 
    &\omega c_{\mathbf{k}}  = \omega_{\mathbf{k}}c_{\mathbf{k}}  + g_{A} \sum_{j} {e}^{- \mathrm{i} \mathbf{k} \cdot \mathbf{r}_{A j}} \psi_{Aj}  + g_{B} \sum_{j} {e}^{- \mathrm{i} \mathbf{k} \cdot \mathbf{r}_{Bj}} \psi_{Bj}. 
    \end{align}  
    \label{eq:eigen eq} 
\end{subequations}
Solving the eigenvalue problem defined by Eqs.~\eqref{eq:eigen eq} yields the band structure of the coupled light-matter system.

\begin{figure}[!htb]
    \centering
    \begin{minipage}[c]{0.48\textwidth}
        \centering
        \includegraphics[width=\textwidth]{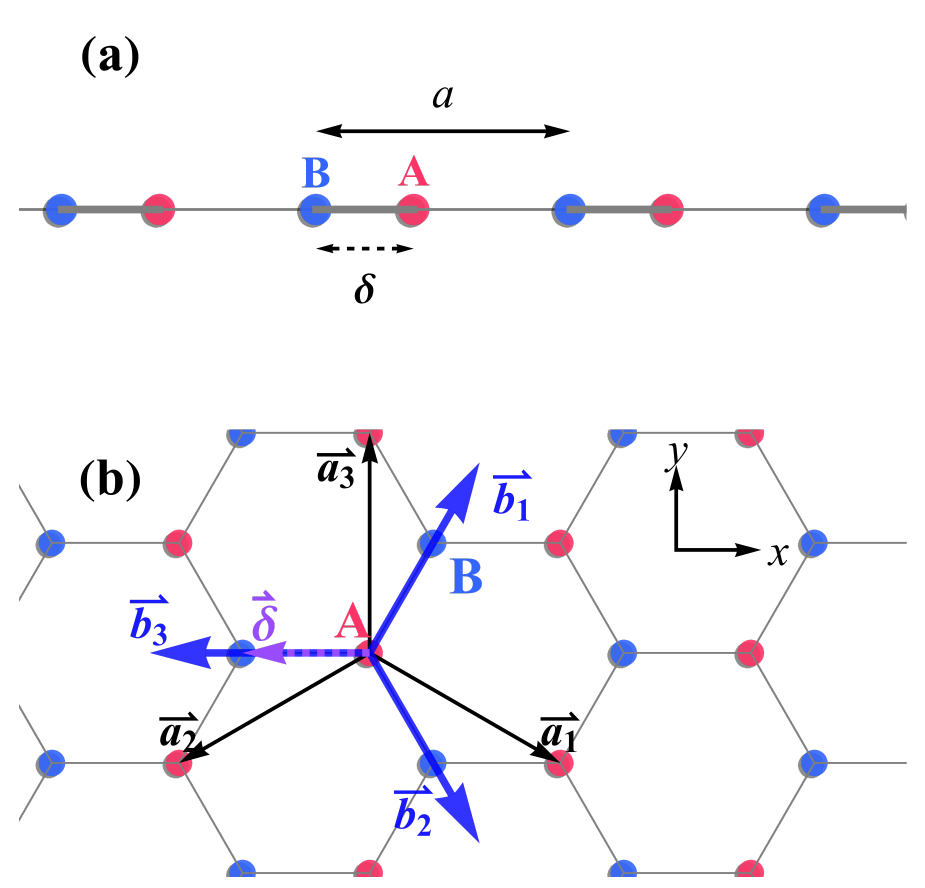}
    \end{minipage}
    \begin{minipage}[c]{0.4\textwidth}
        \centering
        \renewcommand{\arraystretch}{1.1}
 
        \begin{tabular}{||c||c||}
            \hline 
           
            \parbox{2.cm}{\centering\fontsize{10}{0}\textbf{Real Space}} & 
            \parbox{3.5cm}{\centering\fontsize{10}{0}\textbf{Reciprocal Space}} \\
            \hline \hline
            
            $\mathbf{a}_{1} = \left(\frac{\sqrt{3}}{2}, - \frac{1}{2}\right)$ & 
            $\mathbf{b}_{1} = \left(\frac{1}{\sqrt{3}}, 1\right)$ \\
            
            $\mathbf{a}_{2} = \left(-\frac{\sqrt{3}}{2}, - \frac{1}{2}\right)$ & 
            $\mathbf{b}_{2} = \left(\frac{1}{\sqrt{3}}, -1\right)$ \\
            
            $\mathbf{a}_{3} = (0, 1)$ & 
            $\mathbf{b}_{3} = \left(0, -\frac{2}{\sqrt{3}}\right)$ \\
            \hline 
        \end{tabular}
    \end{minipage}

    \vspace{0.5cm} 

    \caption{Lattice geometry and basis definitions. (a) Schematic of the Su-Schrieffer-Heeger (SSH) chain and (b) the 2D honeycomb lattice. The accompanying table summarizes the real-space basis vectors $\mathbf{a}_i$ and the corresponding reciprocal-space basis vectors $\mathbf{b}_j$ used for the honeycomb geometry. The internal displacement between $A$ and $B$ sublattices within a single unit cell is defined by the vector $\bm{\delta} = (-1/\sqrt{3}, 0)$ (purple arrow). Note that the basis vectors satisfy the orthonormality condition $\mathbf{a}_{i} \cdot \mathbf{b}_{j} = \delta_{ij}$.}
    \label{fig:schematicPlot}
\end{figure}

\subsection{\label{subsec:effective hamiltonian}Effective Hamiltonian}
To derive an effective Hamiltonian involving only atomic degrees of freedom, we first define the atomic coordinates. Type-A atoms are located at the points $\mathbf{r}_{Aj} = a \mathbf{R}_{j}$, where $a$ is the lattice constant and $\mathbf{R}_{j} = \sum_{n=1}^{d} j_{n} \mathbf{a}_{n}$ is a dimensionless lattice vector ($j_n \in \mathbb{Z}$), as shown in Fig. \ref{fig:schematicPlot}. Type-B atoms are displaced from the A atoms within the same unit cell at the points $\mathbf{r}_{Bj} = a (\mathbf{R}_{j} + \bm{\delta})$, where $\bm{\delta}$ is the dimensionless displacement vector. Eliminating the amplitudes $c_{\mathbf{k}}$ from Eqs.~\eqref{eq:eigen eq} results in
\begin{subequations} 
\begin{align} 
    &(\omega - \Omega_{A}) \psi_{Aj}  + \frac{g_{A}^{2} V}{c} \sum_{l} G(a \left\vert \mathbf{R}_{jl} \right\vert ;\frac{\omega}{c}) \psi_{Al} + \frac{g_{A} g_{B} V}{c} \sum_{l} G( a \left\vert \mathbf{R}_{jl} + \bm{\delta} \right\vert ;\frac{\omega}{c})  \psi_{Bl} = 0, \\ 
    &(\omega - \Omega_{B}) \psi_{Bj}  + \frac{g_{B}^{2} V}{c} \sum_{l} G(a \left\vert \mathbf{R}_{jl} \right\vert ;\frac{\omega}{c}) \psi_{Bl} + \frac{g_{A} g_{B} V}{c} \sum_{l} G(a \left\vert  \mathbf{R}_{jl} - \bm{\delta} \right\vert ;\frac{\omega}{c}) \psi_{Al} = 0, 
\end{align}
 \label{eq:eigen eq 2} 
\end{subequations}
where $\mathbf{R}_{jl} = \mathbf{R}_{j} - \mathbf{R}_{l}$ is the intercell relative position vector. Here $G$ is the Green's function
\begin{equation} 
    G(\mathbf{r};k) = \int \frac{d^{3} q }{(2\pi)^{3}} \frac{{e}^{\mathrm{i} \mathbf{q} \cdot \mathbf{r}}}{\left\vert \mathbf{q} \right\vert - k }
    =\frac{1}{2\pi^{2} r^{2}} - \frac{\mathrm{i} k}{4 \pi^{2} r} \left[ {e}^{\mathrm{i} k r} E_{1}(\mathrm{i} k r) - {e}^{- \mathrm{i} k r} E_{1}(-\mathrm{i} k r) \right] + \frac{k {e}^{\mathrm{i} k r}}{2 \pi r } , 
    \label{eq:def of green} 
\end{equation}
which introduces an effective long-range interaction between the atoms. We note that $G$ corresponds to the Green's function for the fractional operator $\sqrt{-\Delta} - k$  obeying the radiation condition \cite{kraisler2022collective}. Evidently, $G$ is complex-valued and thus the effective Hamiltonian that results from eliminating the field is non-Hermitian.
%
%

In view of the translational invariance of the lattice, we introduce the Bloch modes by Fourier transforming the atomic amplitudes:
\begin{equation} 
    \psi_{\alpha}(\bm{\beta}) = \sum_{\mathbf{R}_{j}} \psi_{\alpha j} {e}^{-2 \mathrm{i} \pi \bm{\beta} \cdot \mathbf{R}_{j}}, 
    \label{eq:plane wave} 
\end{equation}
where $\bm{\beta} = \sum_{i} \beta_{i} \mathbf{b}_{i}$ is the dimensionless wavevector in the first Brillouin zone (FBZ). Substituting Eq.~\eqref{eq:plane wave} into Eq.~\eqref{eq:eigen eq 2} yields the momentum-space eigenproblem:
\begin{subequations} 
    \begin{align} 
        &\left[ \alpha_{A} - 2 \mathrm{i} \pi \kappa_{A} \alpha^{2} - \kappa_{A} S_{0}(\alpha,\bm{\beta}) \right] \psi_{A}(\bm{\beta})  - \sqrt{\kappa_{A} \kappa_{B}} S_{+}(\alpha,\bm{\beta}) \psi_{B}(\bm{\beta}) = \alpha \psi_{A}(\bm{\beta}) ,\\ 
        &\left[ \alpha_{B} - 2 \mathrm{i} \pi \kappa_{B} \alpha^{2} - \kappa_{B} S_{0}(\alpha,\bm{\beta}) \right] \psi_{B}(\bm{\beta})  -\sqrt{\kappa_{A} \kappa_{B}} S_{-}(\alpha,\bm{\beta}) \psi_{A}(\bm{\beta})= \alpha \psi_{B}(\bm{\beta}),
     \end{align} 
     \label{eq:eigen eq dimless} 
\end{subequations}
where we define the dimensionless energy $\alpha = \omega a /2\pi c$, resonance frequencies $\alpha_{A(B)} = \Omega_{A(B)} a /2\pi c$, and coupling parameters $ \kappa_{A(B)} = g^{2}_{A(B)} V / 2\pi a c^{2}$.

The lattice sums $S_0(\alpha, \bm{\beta})$ and $S_{\pm}(\alpha, \bm{\beta})$ in Eq.~\eqref{eq:eigen eq dimless} describe intra-species and inter-species interactions, respectively:
\begin{subequations} 
    \begin{align} 
        &S_{0}(\alpha,\bm{\beta}) = \sum_{\mathbf{R}_{j} \neq 0} G(\left\vert \mathbf{R}_{j} \right\vert  ; 2\pi \alpha) {e}^{-2 \mathrm{i}  \pi \bm{\beta} \cdot \mathbf{R}_{j}},\\ 
        &S_{\pm}(\alpha,\bm{\beta}) = \sum_{\mathbf{R}_{j}} G(\left\vert \mathbf{R}_{j} \pm \bm{\delta}  \right\vert ;2\pi \alpha) {e}^{- 2 \mathrm{i}  \pi \bm{\beta} \cdot \mathbf{R}_{j} }
        \label{eq:def of lat sum delta}. 
    \end{align} 
    \label{eq:def of lat sum} 
\end{subequations}
In $S_{0}(\alpha,\bm{\beta})$, the divergent $\mathbf{R}_{j}=0$ term is treated separately. The real part of this self-energy term, representing the Lamb shift, is absorbed into the renormalized atomic frequencies $\alpha_{A,B}$. The imaginary part is denoted $2 \pi \kappa_{\alpha} \alpha^{2}$. It accounts for the spontaneous decay rate, which appears in the diagonal term of the Hamiltonian. Detailed derivations of these sums for the 1D lattice are provided in Appendix \ref{appd: 1d lattice sum}, and those for the 2D lattice are given in Appendix~\ref{appd: 2d lattice sum}.

The eigenproblem Eq.~\eqref{eq:eigen eq dimless} is more concisely expressed in terms of $\bm{\psi}(\bm{\beta}) = (\psi_{A}(\bm{\beta}), \psi_{B}(\bm{\beta}))^{{T}}$:
\begin{equation} 
    H(\alpha,\bm{\beta}) \bm{\psi}(\bm{\beta}) = \alpha \bm{\psi}(\bm{\beta}), 
    \label{eq:eigen eq matrix} 
\end{equation}
where the Hamiltonian is decomposed via Pauli matrices as $H(\alpha,\bm{\beta}) = h_{0} \sigma_{0} + \mathbf{h}\cdot \bm{\sigma}$. Here the coefficients are defined by
\begin{subequations} 
    \begin{align} 
        &h_{0}(\alpha,\bm{\beta}) = \frac{1}{2} \left[ \alpha_{A} + \alpha_{B} - 2 \mathrm{i} \pi (\kappa_{A} + \kappa_{B}) \alpha^{2} - (\kappa_{A} + \kappa_{B}) S_{0}(\alpha,\bm{\beta}) \right],\\ 
        &h_{1}(\alpha,\bm{\beta}) =  -\frac{1}{2} \sqrt{\kappa_{A} \kappa_{B}} \left[ S_{+}(\alpha,\bm{\beta}) + S_{-}(\alpha,\bm{\beta}) \right],\\ 
        &h_{2}(\alpha,\bm{\beta}) = \frac{1}{2 \mathrm{i}} \sqrt{\kappa_{A} \kappa_{B}} \left[ S_{+}(\alpha,\bm{\beta}) - S_{-}(\alpha,\bm{\beta}) \right],\\ 
        &h_{3}(\alpha,\bm{\beta}) = \frac{1}{2} \left[ \alpha_{A} - \alpha_{B} - 2 \mathrm{i} \pi (\kappa_{A} - \kappa_{B}) \alpha^{2} - (\kappa_{A} - \kappa_{B}) S_{0}(\alpha,\bm{\beta}) \right]. 
    \end{align} 
    \label{eq:def of h} 
\end{subequations}
These complex coefficients highlight the non-Hermitian nature of the system. Note that $h_0(\alpha, \bm{\beta})$ cannot be eliminated since it is complex-valued, representing an energy shift and decay.

We emphasize that Eq.~\eqref{eq:det eq} defines a nonlinear eigenvalue problem because the Hamiltonian $H(\alpha,\bm{\beta})$ depends on the eigenvalue $\alpha$. The corresponding characteristic equation is given by
\begin{equation} 
    \operatorname{det} \left[ H(\alpha,\bm{\beta}) - \alpha \right] = 0. 
    \label{eq:det eq} 
\end{equation}
Under conditions of weak coupling ($\kappa_{A}, \kappa_{B} \ll 1$) and nearly degenerate atomic frequencies ($\left\vert \alpha_{A} - \alpha_{B} \right\vert \ll \left\vert \alpha_{A} + \alpha_{B} \right\vert$), the nonlinear equation \eqref{eq:det eq} can be solved iteratively. We start with the initial value $\bar{\alpha} = (\alpha_{A} + \alpha_{B})/2 - \mathrm{i} \pi (\kappa_{A} + \kappa_{B})(\alpha_{A} + \alpha_{B})^{2}/4$, which represents the average frequency of two isolated, decaying atoms.
Linearizing the problem by evaluating the Hamiltonian at $\bar{\alpha}$, where $H(\bm{\beta}) := H(\bar{\alpha},\bm{\beta})$, yields two complex eigenvalues:
\begin{equation} 
    \alpha_{\pm}(\bm{\beta}) = h_{0}(\bm{\beta}) \pm \left\vert \mathbf{h}(\bm{\beta}) \right\vert, 
    \label{eq:band structure} 
\end{equation}
where $\left\vert \mathbf{h}(\bm{\beta}) \right\vert = \sqrt{h_{1}^{2}(\bm{\beta}) + h_{2}^{2}(\bm{\beta}) + h_{3}^{2} (\bm{\beta})}$. The real and imaginary parts of $\alpha_{\pm}(\bm{\beta})$ define the dispersion and decay rates, respectively. The  right and left eigenvectors, required for non-Hermitian biorthogonal normalization $\langle \psi_{i}^{L} \lvert \psi_{j}^{R} \rangle = \delta_{ij}$, are:
\begin{subequations} 
\begin{align} 
    &\lvert \psi_{\pm}^{R} (\bm{\beta}) \rangle = \frac{1}{\sqrt{2 \left\vert \mathbf{h}(\bm{\beta}) \right\vert ( \left\vert  \mathbf{h}(\bm{\beta}) \right\vert \mp h_{3}(\bm{\beta}) ) }}(h_{1}(\bm{\beta})- \mathrm{i} h_{2}(\bm{\beta}), \pm \left\vert  \mathbf{h}(\bm{\beta}) \right\vert - h_{3}(\bm{\beta}) )^{\operatorname{T}},\\ 
    &\langle \psi_{\pm}^{L} (\bm{\beta}) \rvert = \frac{1}{\sqrt{2 \left\vert \mathbf{h}(\bm{\beta}) \right\vert ( \left\vert  \mathbf{h}(\bm{\beta}) \right\vert \mp h_{3}(\bm{\beta}) ) }} (h_{1}(\bm{\beta})+ \mathrm{i} h_{2}(\bm{\beta}), \pm \left\vert  \mathbf{h}(\bm{\beta}) \right\vert - h_{3}(\bm{\beta}) ) . 
\end{align}  
\label{eq:eigen states} 
\end{subequations}
These eigenstates will serve as ingredients for characterizing the system's topological invariants in subsequent sections.

\section{\label{sec:SSH}Non-hermitian SSH Model}

\subsection{Band Structure}
\begin{figure}
    \centering
    \includegraphics[width=0.95\textwidth]{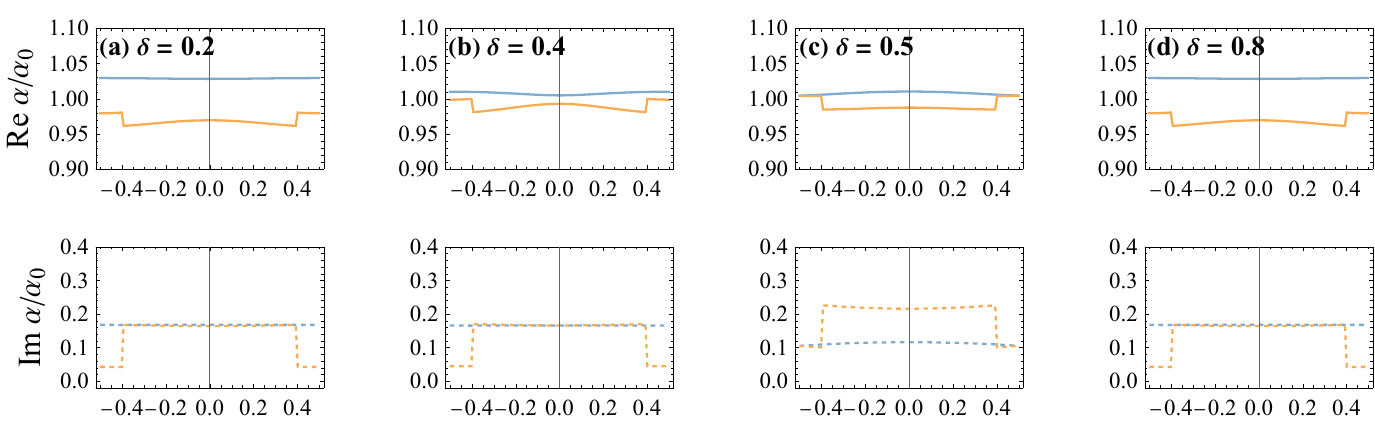}
    \caption{
    Band structure of the effective non-Hermitian SSH model for various values of the intracell separation $\delta$. Throughout this figure (and in all subsequent figures) we fix $\alpha_{A} = \alpha_{B} = 2.4$ and $\kappa_{A} = \kappa_{B} = 0.01$. Panel (a): $\delta = 0.2$; (b): $\delta = 0.4$; (c): $\delta = 0.5$; (d): $\delta = 0.8$. In each subfigure, the upper panel shows the real part of the band structure, while the lower panel shows the corresponding imaginary part.
    }
    \label{fig:ssh band structure}
\end{figure}

In this section, we investigate the band structure, topology, and edge states of a one-dimensional lattice composed of two distinct atom types. We will refer to this as the non-hermitian SSH model, by analogy to its electronic counterpart. We begin by analyzing the band structure of the model. Fig.~ \ref{fig:ssh band structure} displays the complex band structure for varying intracell atomic separations $\delta$. Unless otherwise specified, we fix the parameters $\alpha_{A} = \alpha_{B} = 2.4$ and $\kappa_{A} = \kappa_{B} = 0.01$ for this and all subsequent plots.
%
%
As $\delta$ increases, the band gaps in both the real and imaginary components close at $\delta = 0.5$ and subsequently reopen, indicating a topological phase transition.

A distinct feature of the band structure is the abrupt discontinuity in the imaginary part along momentum lines where $\left\vert \beta \right\vert = (\alpha_{A} + \alpha_{B}) / 2 $. States situated beyond this boundary possess energies below the light line ($\operatorname{Re}~\alpha < |\beta|$), preventing decay via single-photon emission, while conserving energy and momentum. Consequently, these states are subradiant, exhibiting decay rates (imaginary energies) significantly lower than the superradiant states found within the phase boundary.

\subsection{Winding Number}
\begin{figure}
    \centering
    \includegraphics[width=0.95\textwidth]{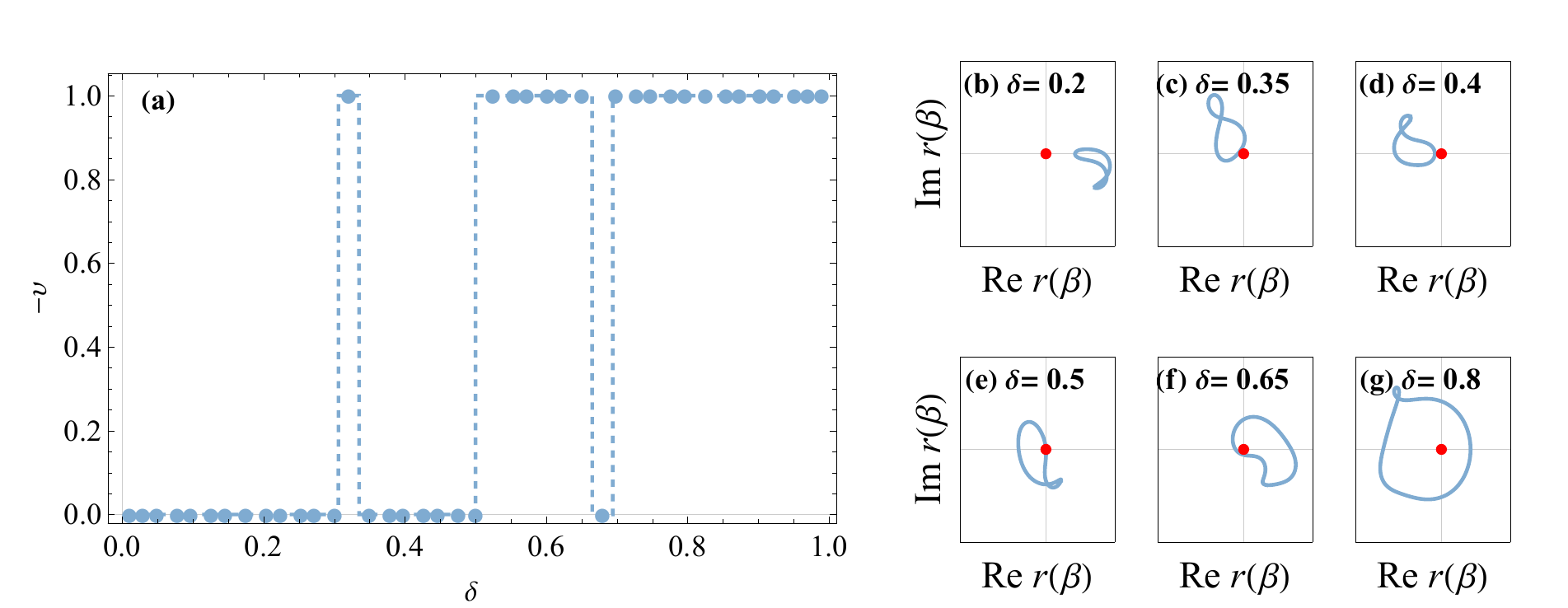}
    \caption{
    Winding number and schematic plots of the complex function $r(\beta) = h_{1}(\beta) + \mathrm{i} h_{2}(\beta)$ in the complex plane. (a) Winding number as a function of $\delta$; (b)–(g) Trajectories of $r(\beta)$ for various values of $\delta$: (b) $\delta = 0.2$; (c) $\delta = 0.35$; (d) $\delta = 0.4$; (e) $\delta = 0.5$; (f) $\delta = 0.65$; (g) $\delta = 0.8$. The red dot indicates the origin in the complex plane. 
    }
    \label{fig:topology}
\end{figure}

%
%
The topological character of this 1D non-Hermitian system is defined by its winding number. Utilizing the left and right eigenvectors from Eq. \eqref{eq:eigen states}, the total winding number for the two bands is given by \cite{shen2018topological}
\begin{equation}
    \upsilon = \upsilon_{+} + \upsilon_{-}  
    = \frac{1}{2\pi \mathrm{i} } \oint d \beta \frac{d}{d \beta} \ln r(\beta),
    \label{eq:winding number}
\end{equation}
where
\begin{align} 
\upsilon_{\pm} = \frac{1}{2\pi} \oint d\beta \langle \psi_{L\pm} \rvert \mathrm{i} \partial_{\beta} \lvert \psi_{R\pm} \rangle, \quad r(\beta)=h_{1}(\beta) + \mathrm{i} h_{2}(\beta). 
\end{align}
Geometrically, this integer invariant represents the number of times the trajectory of $r(\beta)$ encircles the origin in the complex plane as $\beta$ traverses the FBZ.
This point is illustrated in Fig.~\ref{fig:topology}(a), which displays the calculated winding number as a function of $\delta$. The corresponding trajectories of $r(\beta)$ are shown in Figs. \ref{fig:topology}(b)-(g). The winding number is quantized, taking values from $\upsilon=0$ to $\upsilon=1$ as the system undergoes the phase transition at $\delta = 0.5$, where the trajectory intersects the origin. This behavior is physically reasonable: for $\delta < 0.5$, the intracell coupling dominates the intercell hopping, yielding a topologically trivial phase. The opposite limit with $\delta > 0.5$ results in stronger intercell coupling and a non-trivial phase. Notably, the long-range interactions induce additional phase transitions at other values of $\delta$, a phenomenon absent in standard nearest-neighbor SSH models.

\subsection{Dirac Equation and Edge States}
Next we turn to the topic of edge states  and the bulk-edge correspondence. We derive a low-energy continuum model by expanding the Hamiltonian $H(\beta)$ near the Brillouin zone boundary, $\beta_{0}=1/2$. Symmetry constraints require $h_{0}(\beta)$, $h_{1}(\beta)$, and $h_{3}(\beta)$ to be even functions of $q = \beta - \beta_0$, while $h_{2}(\beta)$ is odd. A first-order expansion in $q$ yields
\begin{equation}
    H(\beta_{0}+q) = h_{0} \left(  \beta_{0} \right) \sigma_{0} + h_{1} \left( \beta_{0} \right)\sigma_{x} + q \frac{d}{d\beta} h_{2}(\beta) \Big\rvert_{\beta=\beta_{0}} \sigma_{y} + h_{3}\left( \beta_{0} \right) \sigma_{z} + \mathcal{O}(q^{2}),
    \label{eq:hamiltonian edge}
\end{equation}
Concentrating on states near the vacuum energy, where $\alpha(\beta_{0}) \approx h_{0}(\beta_{0})$, and performing an inverse Fourier transform with respect to $q$, we obtain the real-space Dirac equation for the two-component wavefunction $\bm{\psi}(x)$:
\begin{equation} 
    \left( v_{F} \partial_{x} + m\sigma_{z} + \epsilon \sigma_{x} \right) \bm{\psi}(x) = 0. 
    \label{eq:dirac eq} 
\end{equation}
%
%
Here we define the dimensionless Fermi velocity $v_{F} = h_{2}^{'}(\beta_{0})$, the mass term $m = -h_{1}(\beta_{0})$, and the non-chiral parameter $\epsilon = h_{3}(\beta_{0})$. In this non-Hermitian system, these parameters are generally complex.

\begin{figure}
    \centering
    \includegraphics[width=0.95\textwidth]{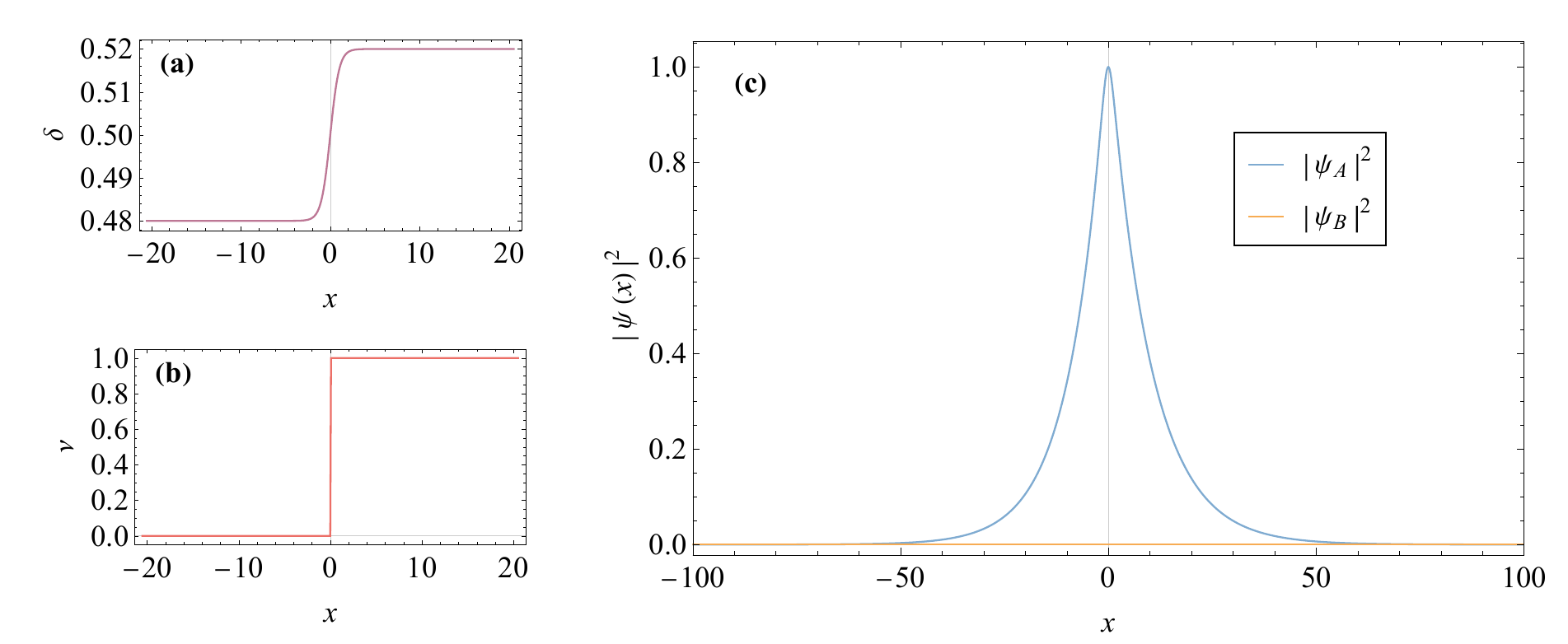}
    \caption{
    Edge state of the effective non-Hermitian SSH model. The intracell atomic separation varies smoothly according to $\delta(x) = 0.02 \tanh(x) + 0.5$. All other parameters are the same as in previous figures. The boundary condition is set as $\psi_{A}(0) = 1$ and $\psi_{B}(0) = 0$. (a) Spatial profile of the intracell separation $\delta(x)$; (b) Corresponding winding number as a function of position; (c) The edge state $\psi_{A}(x)$ is localized near the boundary at $x = 0$, while $\psi_{B}(x)$ remains zero throughout the system.
    }
    \label{fig:edge state}
\end{figure}

When chiral symmetry is preserved ($\alpha_{A} = \alpha_{B}$ and $\kappa_{A} = \kappa_{B}$), implying $\epsilon=0$, the Dirac equation admits an analytical solution for spatially varying mass $m(y)$ and velocity $v_F(y)$:
\begin{equation} 
    \bm{\psi}(x) = \exp{\left(-\int_{0}^{x} dy ~\frac{m(y)}{v_{F}(y)} ~ \sigma_{z}\right)} \bm{\psi}(0). 
    \label{eq: edge state solution}
\end{equation}
To explicitly demonstrate the existence of an edge state, we introduce a domain wall where the intracell separation varies smoothly across the transition point $\delta=0.5$, modeled by $\delta(x) = 0.02 \tanh(x) + 0.5$ (see Fig. \ref{fig:edge state}(a)). This configuration establishes an interface between a topologically trivial region ($\upsilon=0$) and a non-trivial one ($\upsilon=1$) (Fig. \ref{fig:edge state}(b)). Solving Eq. \eqref{eq: edge state solution} under these conditions reveals a solution that is localized at the domain wall ($x=0$), as depicted in Fig. \ref{fig:edge state}(c). The presence of exactly one such solution (edge state), associated with a change in the bulk topological invariant of $\Delta\upsilon=1$, confirms the bulk-edge correspondence in this non-Hermitian setting.

\section{\label{sec:honeycomb}Honeycomb Lattice}

\subsection{Band Structure}
\begin{figure}
   \centering
    \includegraphics[width=0.95\textwidth]{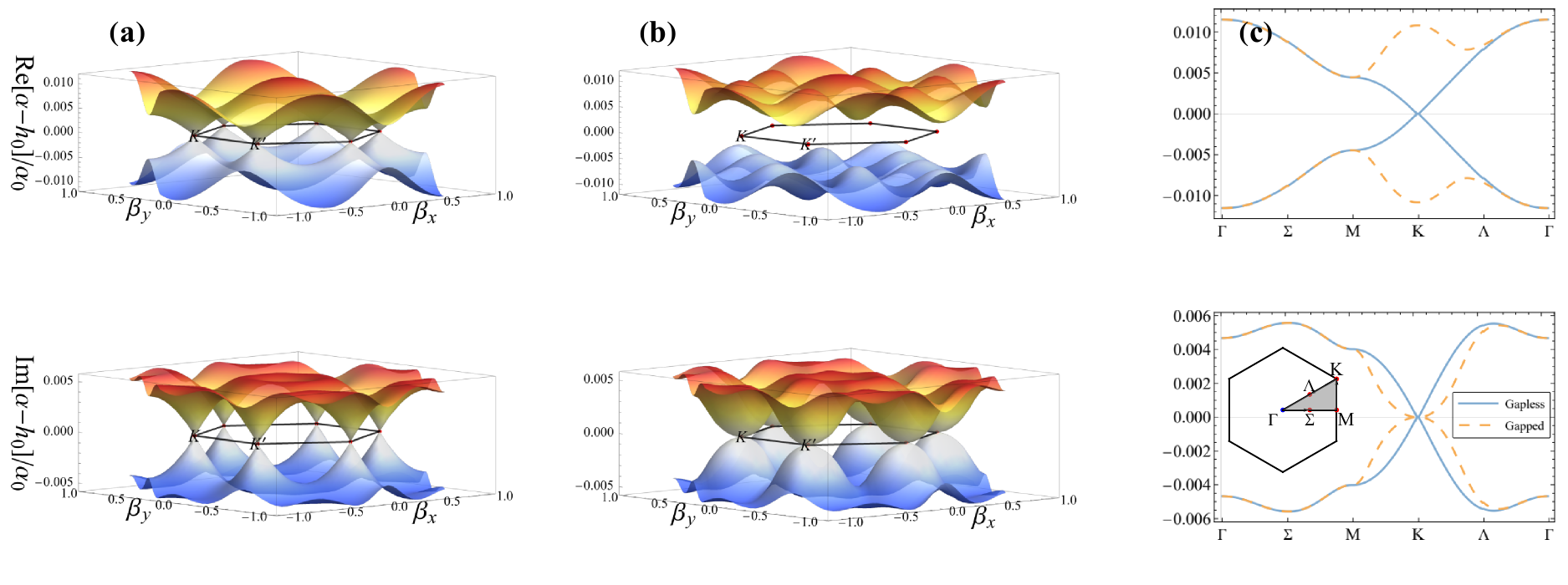}
    \caption{
    Band structure of the non-Hermitian honeycomb lattice. (a) Real (upper panel) and imaginary (lower panel) parts of the band structure without a synthetic gauge field, clearly showing the formation of Dirac cones at the $K$ and $K^{\prime}$ points; (b) band structure with a synthetic gauge field, using $t_{2} = 5 \times 10^{-3}$ and $\phi = \pi / 2$; (c) band structure along the high-symmetry path $\Gamma \rightarrow \Sigma \rightarrow M \rightarrow K \rightarrow \Lambda$ in the Brillouin zone. Solid and dashed lines correspond to cases with and without the gauge field, respectively. In all panels, the band energies are shifted by $h_{0}$, the energy of the vacuum state. 
    }
    \label{fig:honeycomb band structure}
\end{figure}

We now extend our analysis to the two-dimensional honeycomb lattice. Fig.~ \ref{fig:honeycomb band structure} displays the calculated band structure. To elucidate features relative to the band center, we plot the energy difference $\alpha(\bm{\beta}) - h_{0}(\bm{\beta})$. We first examine the symmetric configuration where $\alpha_A = \alpha_B$ and $\kappa_A=\kappa_B$, as shown in Fig. \ref{fig:honeycomb band structure}(a). In this regime, the mass-related term $h_3(\bm{\beta})$ vanishes, ensuring the system preserves chiral symmetry. Furthermore, the system exhibits time-reversal symmetry, satisfying $H(\bm{\beta}) = H(-\bm{\beta})^{{T}}$. These combined symmetries protect the gapless points in the band structure—the Dirac points—located at the corners of the FBZ ($K$ and $K^{\prime}$). In the vicinity of these points, the bands form Dirac cones characterized by the linear dispersion relation $\alpha(\bm{\beta}_{K(K^{\prime})}+\mathbf{q}) = \pm v_{F} \left\vert \mathbf{q} \right\vert$, a property we derive in Appendix \ref{appd:symmetry of lat sum}. As in the 1D case, the Fermi velocity $v_F$ is complex.

To produce a non-trivial topological phase, a band gap must be opened. This can be accomplished by breaking either chiral symmetry (setting $\alpha_A \neq \alpha_B$) or time-reversal symmetry. While breaking chiral symmetry is experimentally straightforward, it yields a topologically trivial phase. Instead, we focus on breaking time-reversal symmetry. Unlike charged electrons, neutral atoms cannot couple directly to a magnetic field; therefore, we employ a synthetic gauge field \cite{Dalibard2011, Jotzu2014,Cooper2019}. Adopting Haldane's approach, we introduce a complex next-nearest-neighbor hopping term in the Hamiltonian. Accordingly, we modify $h_0$ and $h_3$ as
\begin{subequations} 
    \begin{align} 
        h_{0}(\bm{\beta}) \to  h_{0}(\bm{\beta}) + 2t_{2} \cos \phi \sum_{i=1}^{3} \cos(2 \pi \bm{\beta} \cdot \mathbf{a}_{i}),\\ 
        h_{3}(\bm{\beta}) \to  h_{3}(\bm{\beta}) - 2t_{2} \sin \phi \sum_{i=1}^{3} \sin(2 \pi \bm{\beta} \cdot \mathbf{a}_{i}).
     \end{align} \label{eq:Haldane term}
\end{subequations}
where $t_2$ is the hopping amplitude and $\phi$ the phase of the artificial gauge field. As illustrated in Fig.~\ref{fig:honeycomb band structure}(b), this term opens a gap in the real part of the energy spectrum at the Dirac points. The gap in the imaginary part remains closed because the additional term in $h_3(\bm{\beta})$ is purely real and vanishes at the $K$ and $K^{\prime}$ Dirac points. 
%
%
Fig.~\ref{fig:honeycomb band structure}(c) presents a direct comparison of the band structures along high-symmetry paths. The primary modification occurs near the Dirac points, where the imaginary component of the dispersion relation transitions from linear to parabolic.

\subsection{Chern Number}
For a non-Hermitian system with distinct left ($L$) and right ($R$) eigenvectors, we define the Berry curvature as
\begin{equation}
    B_{n,\mu \nu}^{i j}(\bm{\beta}) = \mathrm{i} \langle \partial_{\mu} \psi_{n}^{i} (\bm{\beta}) \vert  \partial_{\nu} \psi_{n}^{j} (\bm{\beta}) \rangle
\end{equation}
where $i,j  = L, R$. Integrating the curvature over the FBZ yields the Chern numbers $C_n^{ij}$ \cite{shen2018topological}:
\begin{equation}
    C_{n}^{ij } = \frac{1}{2 \pi } \int_{\text{FBZ}} d \beta^{\mu} \wedge d \beta^{\nu} B_{n,\mu \nu}^{i j}(\bm{\beta})
    \label{eq:def of Chern number}
\end{equation}
In view of the relations $C_{n}^{LR}= C_{n}^{RL*}$ and $\sum_{n} C_{n}^{ij} = 0$, it suffices to consider the left-right Chern number of the lower band. Henceforth, we omit the band index and denote this quantity as $C^{LR}$. 

\begin{figure}
    \centering
    \includegraphics[width=0.8\textwidth]{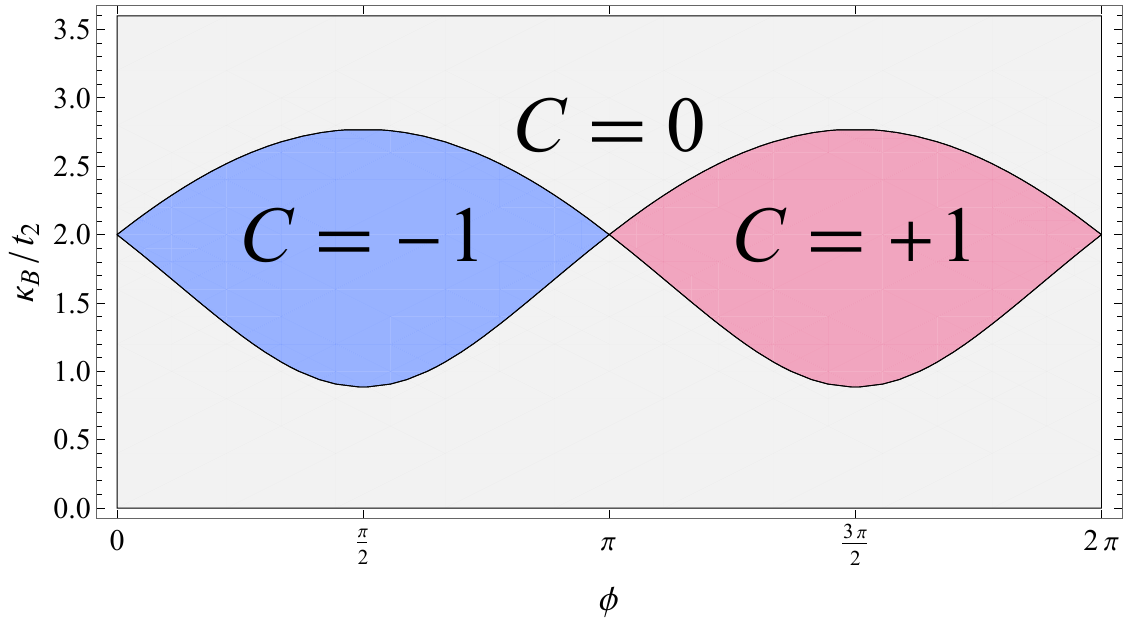}
    \caption{
    Phase diagram of the honeycomb lattice. The parameters are fixed as $\alpha_{A} = \alpha_{B} = 2.4$, $\kappa_{A} = 0.01$, and $t_{2} = 5 \times 10^{-3}$, while $\kappa_{B}$ (the coupling strength of the $B$-type atom) and the phase $\phi$ of the next-nearest-neighbor hopping are varied. The Chern number is $+1$ in the red region, $-1$ in the blue region, and $0$ in the gray region.
    }
    \label{fig:phase diagram}
\end{figure}

The synthetic gauge field breaks time-reversal symmetry and introduces a staggered mass term. It follows from Eq.~\eqref{eq:Haldane term} that the effective mass term at the $K$ and $K^{\prime}$ points are \cite{Haldane1988,li2022gain}:
\begin{equation}
    m_{K} = h_{3}(\bm{\beta}_{K}) + 3\sqrt{3} t_{2} \sin \phi, \quad m_{K^{\prime}} = h_{3}(\bm{\beta}_{K}) - 3\sqrt{3} t_{2} \sin \phi.
    \label{eq:mass term at K and K'}
\end{equation} 
For the low-energy Dirac model, the total Chern number is
\begin{equation}
    C = \frac{1}{2}(C^{LR}+C^{RL}) = \frac{1}{2} \operatorname{Re}\left[ m_{K^{\prime}} / \sqrt{ m_{K^{\prime}}^{2} } - m_{K} /  \sqrt{m_{K}^{2}}   \right].
    \label{eq:chern number}
\end{equation}
which is quantized to 0 or $\pm 1$. The resulting topological phase diagram is depicted in Fig. \ref{fig:phase diagram}. 
%

\subsection{Dirac Equation and Edge States}
We now turn to the derivation of a continuum model for the honeycomb lattice. To proceed, we expand the Hamiltonian near the Dirac points. As described in Appendix \ref{appd:symmetry of lat sum}, lattice symmetries impose strong constraints on this expansion. A key result is that the expansion is isotropic to first order in the momentum deviation $\mathbf{q}$, defined by a single complex Fermi velocity $v_F$. The idea applies generally: provided that the interaction preserves the point group symmetry ($C_{3v}$), the Fermi velocity remains isotropic. Consequently, the Hamiltonian near the $K$ and $K^{\prime}$ points takes the form:
\begin{align}
    H(\bm{\beta}_{K} +\mathbf{q}) &=  \alpha_{0} \sigma_{0} - v_{F} q_{y} \sigma_{x}+ v_{F} q_{x} \sigma_{y}+ m_{K} \sigma_{z} + \mathcal{O}(\mathbf{q}^{2}),\\
     H(\bm{\beta}_{K^{\prime}} +\mathbf{q}) &= \alpha_{0} \sigma_{0} + v_{F} q_{y} \sigma_{x}+ v_{F} q_{x} \sigma_{y}+ m_{K^{\prime}} \sigma_{z} + \mathcal{O}(\mathbf{q}^{2}).
     \label{eq:Dirac equations}
\end{align}
Here $\alpha_0 = h_0(\bm{\beta}_K) =h_0(\bm{\beta}_{K^{\prime}}) $, $m_{K(K')}$ are the mass terms defined in Eq.~\eqref{eq:mass term at K and K'}, and the momentum is expanded in the orthonormal basis $\{\mathbf{b}_{x} = (1,0),\mathbf{b}_{y} = (0,1)\}$ as $\mathbf{q} = q_{x} \mathbf{b}_{x} + q_{y} \mathbf{b}_{y}$. Applying the inverse Fourier transform ($\mathbf{q} \to -\mathrm{i}\nabla$) yields Dirac equations governing the real-space dynamics of the wave packet envelope $\bm{\psi}(x,y,\tau)$:
\begin{align}
   &(\alpha_{0} \sigma_{0} + \mathrm{i} v_{F} \partial_{y} \sigma_{x} -   \mathrm{i} v_{F} \partial_{x} \sigma_{y} + m_{K} \sigma_{z}) \bm{\psi} (x,y,\tau) = \mathrm{i} \partial_{\tau} \bm{\psi}(x,y,\tau), \label{eq:Dirac eq K} \\ 
   &(\alpha_{0} \sigma_{0} -\mathrm{i} v_{F} \partial_{y} \sigma_{x} -    \mathrm{i} v_{F} \partial_{x} \sigma_{y} + m_{K^{\prime}} \sigma_{z}) \bm{\psi} (x,y,\tau) = \mathrm{i} \partial_{\tau}  \bm{\psi} (x,y,\tau).\label{eq:Dirac eq Kp}
\end{align}
where $\tau = 2 \pi c t / a$ is the dimensionless time and $a$ is the lattice constant.

\subsection{Edge State Dynamics}
We now discuss the dynamics of chiral edge states at a domain wall. Here we assume that the mass term $m(y)$ changes its sign. Focusing on the $K$ valley, we employ an ansatz that is a plane wave in $x$ and an undetermined function $f(y)$ in $y$:
\begin{equation}
    \psi_{\pm}(x,y,\tau) = \mathcal{N} {e}^{\mathrm{i} q_{x} x} {e}^{-\mathrm{i} \alpha_{\pm}(q_{x}) \tau} f(y) \begin{pmatrix}
        1 \\
        \pm \mathrm{i}
    \end{pmatrix},
    \label{eq:ansatz}
\end{equation}
where $\mathcal{N}$ is a normalization constant and $\alpha_{\pm}(q_{x})$ is the band energy. Given the linear dispersion near the Dirac points, we have $\alpha_{\pm}(q_{x}) = \alpha_{0} \pm v_{F} q_{x}$. Substituting Eq. \eqref{eq:ansatz} into Eq. \eqref{eq:Dirac eq K} yields 
\begin{equation}
    v_{F}\partial_{y} f(y) =  \pm m(y) f(y),
\end{equation} 
which has the solution $f(y) \propto \exp(\pm \int dy' m(y')/v_{F})$. Localization at the domain wall ($y=0$) requires the wavefunction to decay away from the interface, rendering only one sign of the exponent physically admissible. This ensures the existence of a single chiral mode at the interface, a manifestation of the bulk-edge correspondence where $\Delta C=1$.

\begin{figure}
   \centering
    \includegraphics[width=0.95\textwidth]{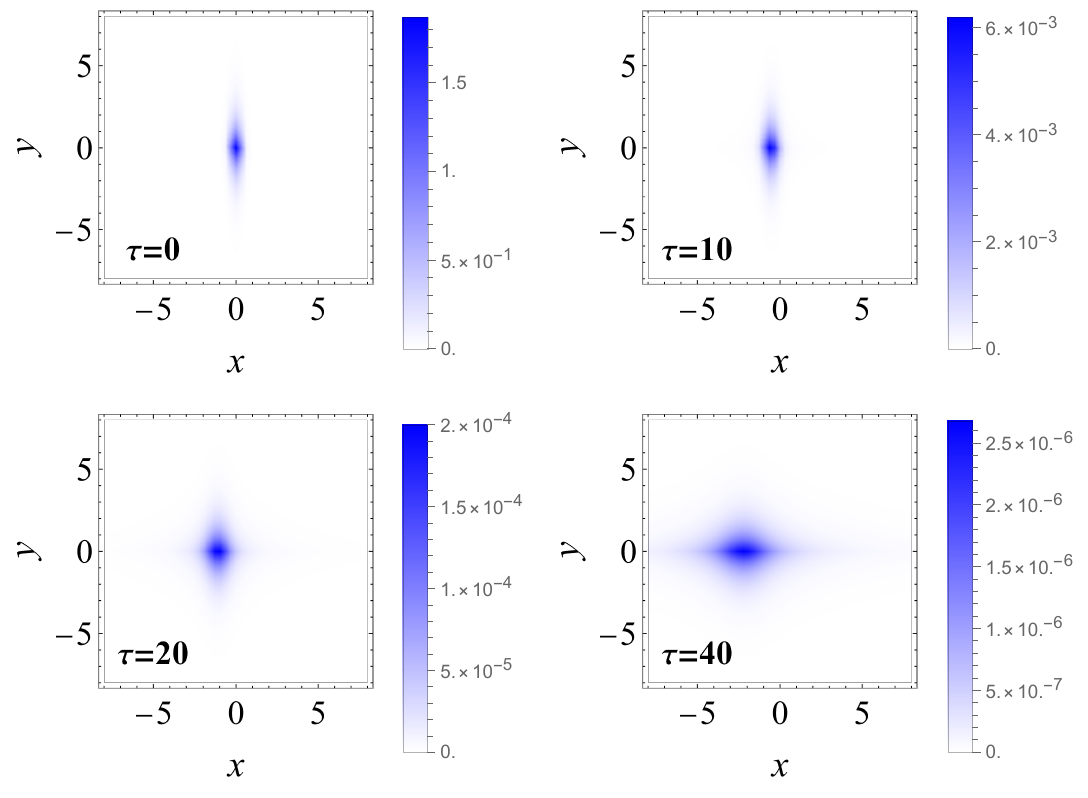}
    \caption{
    Dynamics of the edge state in the honeycomb lattice. The mass term is taken as $m(y) = t_{2} \tanh(y / y_{0})$, with $t_{2} = 5 \times 10^{-3}$ and $y_{0} = 0.1$, so the domain wall is along the $y=0$. The $x$ dependence of the initial wavefunction is chosen as $\psi(x, y, 0) =\mathcal{N} {e}^{-x^{2} / x_{0}^{2}}  {e}^{-\int_{0}^{y} dz ~ m(z) /v_{F}}$ for $x_{0} = 0.5$.
    }
    \label{fig:edge state honeycomb}
\end{figure}

A general localized solution can be constructed as a wave packet by superposing the $\psi_{-}$ basis states from Eq.~\eqref{eq:ansatz} (the $\psi_{+}$ solutions are exponentially growing and can be discarded). Given an initial Gaussian wave packet localized in $x$ of the form  ${e}^{-x^{2} / x_{0}^{2}}$, the time evolution is given by
\begin{equation} \psi(x,y,\tau) = \mathcal{N} {e}^{-\int_{0}^{y} dz ~ m(z) /v_{F}} {e}^{-\mathrm{i} \alpha_{0} \tau}\int d q_{x} ~{e}^{\mathrm{i} q_{x} (x +v_{F} \tau) - q_{x}^{2} x_{0}^{2} / 4 } \begin{pmatrix} 1 \\ \mathrm{- i} \end{pmatrix}. \end{equation}
The solution is plotted in Fig. \ref{fig:edge state honeycomb} for the domain wall $m(y) = t_{2} \tanh(y / y_0)$. We see that the wave packet remains localized at the interface ($y=0$) and propagates chirally along the negative $x$ direction. The decay of the amplitude in time is a signature of the system's non-Hermitian nature.

\section{\label{sec:discuss}Discussion}
We have established a quantum-optical framework for calculating the band structure and topological properties of atomic lattices governed by long-range, radiative interactions. Our approach begins with the derivation of an effective non-Hermitian two-band Hamiltonian, which governs the system's band structure and dynamics within the single-excitation subspace.

A central result of this study is the demonstration of the existence of Dirac points and the emergence of effective Dirac dynamics near resonance. We emphasize that this finding is not limited to the specific atomic arrays considered here, but is a general feature of systems preserving the underlying lattice symmetry. Our proof reveals that the formation of Dirac points and the validity of the Dirac equation with a uniform Fermi velocity is dictated by symmetry constraints, which is independent of the details of the coupling. As a consequence, our results are applicable in diverse settings even when the fields modes are spatially confined, such as in waveguide QED. As long as the lattice symmetry remains unbroken, the effective Dirac physics persists, regardless of whether the interactions are short- or long-ranged and Hermitian or non-Hermitian.

We analyzed the complex band structures of the 1D SSH and 2D honeycomb models and observed the formation of Dirac cones. By deriving the corresponding continuum models near the Dirac points, we obtained Dirac equations characterized by an isotropic Fermi velocity. In contrast to the Hermitian case, the Fermi velocity becomes complex-valued, a feature that is linked to the non-Hermitian skin effect. Furthermore, we investigated the topological phases of these systems by calculating their topological invariants and  edge states. The relation between the bulk invariant and the number of localized states at a domain wall explicitly verifies the bulk-edge correspondence in this non-Hermitian setting.

We conclude by highlighting several promising avenues for future research. First, our current study focuses on single-excitation states. However, exploring the many-body physics of the system when multiple excitations are present remains an intriguing open question \cite{MorenoCardoner2022, Mahmoodian2018}. Second, while we assumed the atoms are fixed at lattice sites, experiments are performed under conditions when atomic motion is present. Investigating the influence of this motion on the band structure and topological properties of the system would be of interest \cite{Shahmoon2020}. Finally, we restricted our attention to two-band models. The extension to multi-band models, as arises when considering atoms with more than two energy levels or unit cells containing multiple atoms is another topic for further study \cite{Masson2020, Parmee2020}.

\section*{ACKNOWLEDGMENTS}
We thank P. Wang, P. de Maat, and T. Hong for helpful discussions. 

\section*{Data availability}
The data and codes are available from the authors upon reasonable request.


\appendix

\section{\label{appd: 1d lattice sum}One-dimensional Lattice Sum}
In this section, we evaluate the one-dimensional lattice sum $S_{\pm}(\alpha,\beta)$, defined as
\begin{equation}
    S_{\pm}(\alpha,\bm{\beta}) = \sum_{\mathbf{R}_{j}} G(\left\vert \mathbf{R}_{j} \pm \bm{\delta} \right\vert ; 2\pi \alpha) {e}^{-2 \mathrm{i} \pi \bm{\beta} \cdot \mathbf{R}_{j}},
\end{equation}
where the Green's function is given by
\begin{equation}
    G(\mathbf{r};k) = \frac{1}{2\pi^{2} r^{2}} - \frac{\mathrm{i} k}{4 \pi^{2} r} \left[ {e}^{\mathrm{i} k r} E_{1}(\mathrm{i} k r) - {e}^{-\mathrm{i} k r} E_{1}(-\mathrm{i} k r) \right] + \frac{k {e}^{\mathrm{i} k r}}{2 \pi r }. 
\end{equation} 
Here $\mathbf{R}_{j}$ denotes the lattice vectors, $\bm{\delta}$ is the displacement vector between the $A$ and $B$ sublattices, and $\bm{\beta}$ is the crystal momentum confined to the FBZ. 

We decompose the sum into three terms:
\begin{equation}
   S_{\pm}(\alpha,\beta) = S_{\pm1}(\alpha,\beta) + S_{\pm2}(\alpha,\beta) + S_{\pm3}(\alpha,\beta).
   \label{eq:decomposition of lat sum}
\end{equation}
These components are defined as follows:
\begin{subequations}
    \begin{align}
    &S_{\pm1}(\alpha,\beta) = \frac{1}{2\pi^{2}} \sum_{n \in \mathbb{Z}} \frac{{e}^{-2 \mathrm{i} \pi \beta n}}{\left\vert n \pm \delta \right\vert^{2}}, \\
    &S_{\pm2}(\alpha,\beta) = \frac{\alpha}{2\pi \mathrm{i}} \sum_{n \in \mathbb{Z}} \frac{{e}^{-2 \mathrm{i} \pi \beta n}}{\left\vert n \pm \delta \right\vert} \left[ {e}^{2 \mathrm{i} \pi \alpha \left\vert n \pm \delta \right\vert} E_{1}(2 \mathrm{i} \pi \alpha \left\vert n \pm \delta \right\vert) - (\alpha \to -\alpha) \right], \\
    &S_{\pm3}(\alpha,\beta) = \alpha \sum_{n \in \mathbb{Z}} \frac{{e}^{2 \mathrm{i} \pi \alpha \left\vert n \pm \delta \right\vert}}{\left\vert n \pm \delta \right\vert} {e}^{-2 \mathrm{i} \pi \beta n}.
    \end{align}
    \label{eq:lat sum delta appx}
\end{subequations}
In the following, we focus on computing the sums with the ``$+$'' sign. Results for the ``$-$'' sign are obtained by the substitution $\delta \to -\delta$.

\subsection{\label{subappx:S1} $S_{1+}$ in 1D}
A convenient method to evaluate the lattice sum is the Poisson summation formula. We begin by computing the Fourier transform of the summand in $S_{1+}(\alpha,\beta)$:
\begin{equation}
    \frac{1}{2\pi^{2}}\int_{-\infty}^{\infty} dx \frac{{e}^{-2 \mathrm{i} \pi f x }}{ (x+\delta)^{2}}  = - \left\vert  f \right\vert {e}^{2 \mathrm{i} \delta f}.
\end{equation}
Applying the Poisson summation formula, we obtain:
\begin{equation}
    S_{1+}(\alpha,\beta) = - {e}^{2 \mathrm{i} \pi q} \sum_{k \in \mathbb{Z}} \left\vert  k + \beta \right\vert {e}^{2 \mathrm{i} \pi \delta k }.
\end{equation}
The series above is computed by analytic continuation. 
%
%
This yields
\begin{equation}
    S_{1+}(\alpha,\beta) = -{e}^{2 \mathrm{i} \pi q} \left[ \left\vert \beta \right\vert  + \beta \frac{1+ {e}^{ 2 \mathrm{i} \pi \delta}}{ 1 - {e}^{ 2 \mathrm{i} \pi \delta}} + \frac{2 {e}^{ 2 \mathrm{i} \pi \delta}}{ (1- {e}^{2 \mathrm{i} \pi \delta})^{2}} \right].
    \label{eq:S1 0}
\end{equation}
Equation \eqref{eq:S1 0} is further simplified resulting in
\begin{equation}
    S_{1+}(\alpha,\beta) = {e}^{2 \mathrm{i} \pi q} \left[ 1 /2 +  \cot^{2}(\pi \delta) / 2 - \left\vert \beta \right\vert - \mathrm{i} \beta \cot(\pi \delta) \right].
    \label{eq:S1}
\end{equation}
Other approaches, utilizing complex integration, can also be used to evaluate the lattice sum $S_{1+}(\alpha,\beta)$ and yield the same result as Eq.~\eqref{eq:S1}.

\subsection{\label{subappx:S2} $S_{2+}$ in 1D}
Following the same approach used for $S_{1+}(\alpha,\beta)$, we apply the Poisson summation formula to evaluate $S_{2+}(\alpha,\beta)$. We begin by computing the Fourier transform of its summand:
\begin{equation}
    \int_{-\infty}^{\infty} dx \frac{{e}^{-2 \mathrm{i} \pi  f x }}{2 \pi \mathrm{i} \left\vert x \right\vert } \left[ {e}^{2 \mathrm{i} \pi \alpha \left\vert  x \right\vert } E_{1}(2 \mathrm{i} \pi \alpha \left\vert  x \right\vert ) - {e}^{-2 \mathrm{i} \pi \alpha \left\vert  x \right\vert } E_{1}(-2 \mathrm{i} \pi \alpha \left\vert  x \right\vert )\right] =    \gamma_{\text{E}}+ \ln( \alpha + \left\vert  f \right\vert ),
\end{equation}
where $\gamma_{\text{E}}$ denotes the Euler constant. Applying the Poisson summation formula yields
\begin{equation}
    S_{2+}(\alpha,\beta) = \alpha {e}^{ 2 \mathrm{i} \pi q} \left[ \gamma_{\text{E}}\sum_{k \in \mathbb{Z}} {e}^{2 \mathrm{i} \pi \delta k} + \sum_{k \in  \mathbb{Z}} \ln(\alpha + \left\vert  \beta + k \right\vert ) {e}^{ 2 \mathrm{i} \pi \delta k}\right].
    \label{eq:S2 0}
\end{equation}
The first sum vanishes for non-integer $\delta$. The second sum can be expressed in terms of the Lerch transcendent $\Phi(z,s,a) = \sum_{n=0}^{\infty} z^{n} / (n+a)^{s}$, using the identity
\begin{equation}
    \sum_{k=1}^{\infty} \ln(x+k) {e}^{2 \mathrm{i} \pi \delta k} = - {e}^{ 2 \mathrm{i} \pi \delta } \partial_{2} \Phi({e}^{2 \mathrm{i} \pi \delta},0,x+1),
\end{equation}
which can be suitably analytically continued. 
%
%
Here the partial derivative is taken with respect to the second argument of the Lerch transcendent. Substituting this identity into Eq.~\eqref{eq:S2 0}, we obtain
\begin{equation}
    S_{2+}(\alpha,\beta) = \alpha {e}^{ 2 \mathrm{i} \pi q} \ln(\alpha + \left\vert  \beta \right\vert ) - \alpha {e}^{ 2 \mathrm{i} \pi q} \left[ {e}^{ 2 \mathrm{i} \pi \delta }  \partial_{2} \Phi({e}^{2 \mathrm{i} \pi \delta},0,1+\alpha+\beta)  + ( \delta, \beta ) \to  (-\delta, -\beta)\right].
    \label{eq:S2}
\end{equation}

\subsection{\label{subappx:S3} $S_{3+}$ in 1D}
We compute $S_{3+}$ directly by breaking the sum into parts:
\begin{equation}
    S_{3+}(\alpha,\beta) = \alpha \left[ \frac{{e}^{2 \mathrm{i} \pi \alpha \left\vert  \delta \right\vert }}{\left\vert \delta \right\vert } + {e}^{ 2 \mathrm{i} \pi \delta \alpha } \sum_{n=1}^{\infty} \frac{ {e}^{ 2 \mathrm{i} \pi (\alpha - \beta) n }}{ n + \delta} +  {e}^{ - 2 \mathrm{i} \pi \delta \alpha } \sum_{n=1}^{\infty} \frac{ {e}^{ 2 \mathrm{i} \pi (\alpha + \beta) n }}{ n - \delta} \right].
\end{equation}
Using the definition of the Lerch transcendent function, we have
\begin{equation}
      S_{3+}(\alpha,\beta) = \alpha \frac{{e}^{2 \mathrm{i} \pi \alpha  \left\vert \delta \right\vert   } }{ \left\vert \delta \right\vert } + \alpha \left[ {e}^{2 \mathrm{i} \pi ( \alpha (1+ \delta) - \beta )} \Phi({e}^{ 2 \mathrm{i} \pi (\alpha - \beta)},1,1+\delta) +  ( \delta, \beta ) \to  (-\delta, -\beta)  \right].
      \label{eq:S3} 
\end{equation}

\subsection{Result of 1D Lattice Sum}
Finally, combining the results from Eqs.~\eqref{eq:S1}, \eqref{eq:S2}, and \eqref{eq:S3}, we obtain the closed form expression for the lattice sum $S_{+}(\alpha,\beta)$ by using the decomposition in Eq.~\eqref{eq:decomposition of lat sum}:
\begin{align}
   &S_{+}(\alpha,\beta) \nonumber \\
   &= {e}^{ 2 \mathrm{i} \pi q} \left[ 1 /2 +  \cot^{2}(\pi \delta) / 2 - \left\vert \beta \right\vert - \mathrm{i} \beta \cot( \pi \delta) \right] + \alpha {e}^{2\mathrm{i} \pi q} \ln(\alpha + \left\vert \beta \right\vert ) +\alpha {e}^{2 \mathrm{i} \pi \alpha \delta} / \delta \nonumber \\
   &-\alpha {e}^{ 2 \mathrm{i} \pi q} \left[ {e}^{  2 \mathrm{i} \pi  \delta} \partial_{2} \Phi({e}^{ 2 \mathrm{i} \pi  \delta},0,1+\alpha+\beta) + {e}^{ - 2 \mathrm{i} \pi \delta} \partial_{2} \Phi({e}^{- 2 \mathrm{i} \pi \delta},0,1+\alpha-\beta) \right] \nonumber \\
   & + \alpha {e}^{2 \mathrm{i} \pi \alpha} \left[ {e}^{2 \mathrm{i} \pi ( \delta \alpha - \beta )} \Phi({e}^{ 2 \mathrm{i} \pi (\alpha - \beta)},1,1+\delta) + {e}^{-2 \mathrm{i} \pi ( \delta \alpha - \beta )} \Phi({e}^{ 2 \mathrm{i} \pi (\alpha + \beta)},1,1-\delta)\right].
   \label{eq:lattice sum delta}
\end{align}
The expression for $S_{0}$ is derived by subtracting the divergent $n=0$ term from Eq.~\eqref{eq:lattice sum delta} and then taking the limit $\delta \to 0$. This yields
\begin{align}
   S_{0}(\alpha,\beta) =& {B}_{2}(\left\vert \beta \right\vert)  -2\alpha^{2}+2\alpha^{2} \ln \alpha + \alpha \ln (2 \pi (\alpha+\left\vert \beta \right\vert ))-\alpha \ln \Gamma(1+\alpha+  \beta )\nonumber \\
    &-\alpha \ln \Gamma(1+\alpha-  \beta )-\alpha \ln (1-{e}^{ 2 \mathrm{i}  \pi (\alpha +  \beta  )})-\alpha \ln (1-{e}^{ 2 \mathrm{i}  \pi (\alpha -  \beta  )}),
    \label{eq:lattice sum 0}
\end{align}
where ${B}_{2}(x)$ is the second-order Bernoulli polynomial and $\ln \Gamma(x)$ is the log-gamma function. We have numerically verified Eqs.~\eqref{eq:lattice sum delta} and \eqref{eq:lattice sum 0}.

\section{\label{appd: 2d lattice sum}Two-dimensional Lattice Sum}
As in the one-dimensional case, we decompose the two-dimensional lattice sum into three terms:
\begin{subequations}
\begin{align}
    &S_{\pm1}(\alpha,\bm{\beta}) = \frac{1}{2 \pi^{2}} \sum\limits_{\mathbf{R}} \frac{{e}^{-2 \mathrm{i} \pi \bm{\beta} \cdot \mathbf{R} }}{ \left( \mathbf{R} \pm \bm{\delta} \right)^{2} } \\
    &S_{\pm2}(\alpha,\bm{\beta}) = \frac{ \alpha}{2 \pi \mathrm{i} } \sum_{\mathbf{R}} \frac{ {e}^{- 2 \mathrm{i} \pi \bm{\beta} \cdot \mathbf{R} }}{ \left\vert  \mathbf{R}  \pm \bm{\delta} \right\vert } \left[ {e}^{2 \mathrm{i} \pi \alpha \left\vert  \mathbf{R}  \pm \bm{\delta} \right\vert  } E_{1}(2 \mathrm{i} \pi \alpha \left\vert   \mathbf{R}  \pm \bm{\delta} \right\vert ) - (\alpha \to  - \alpha)\right], \\
    &S_{\pm3}(\alpha,\bm{\beta}) = \alpha \sum_{\mathbf{R}} \frac{{e}^{ 2 \mathrm{i} \pi \alpha \left\vert  \mathbf{R} \pm \bm{\delta} \right\vert }}{ \left\vert  \mathbf{R} \pm \bm{\delta} \right\vert } {e}^{-2 \mathrm{i} \pi \bm{\beta} \cdot \mathbf{R} }.
\end{align}
\label{eq:def of lat sum 2d}
\end{subequations}
Here the lattice points are defined as $\mathbf{R} = j_{1} \mathbf{a}_{1} + j_{2} \mathbf{a}_{2}$ with $(j_{1}, j_{2} \in \mathbb{Z})$, forming a hexagonal lattice. The basis vectors of the lattice are given by $\mathbf{a}_{1} = \left( \sqrt{3} / 2, -1 / 2 \right)$ and $\mathbf{a}_{2} = \left( -\sqrt{3} / 2, -1 / 2 \right)$. The crystal momentum is defined as $\bm{\beta} = \beta_{1} \mathbf{b}_{1} + \beta_{2} \mathbf{b}_{2}$, constrained within the FBZ. The reciprocal lattice basis vectors are given by $\mathbf{b}_{1} = \left( 1 / \sqrt{3}, 1 \right)$ and $\mathbf{b}_{2} = \left( 1 / \sqrt{3}, -1 \right)$. The vector $\bm{\delta} = \delta_{1} \mathbf{a}_{1} + \delta_{2} \mathbf{a}_{2}$ represents the displacement between the $A$ and $B$ sublattice atoms within the same unit cell.

\subsection{$S_{1+}$ in 2D}
We begin by noting the following integral identity
\begin{equation}
    \int_{0}^{1} dt~ t^{x - 1} = \frac{1}{x}.
    \label{eq:int 1}
\end{equation}
Substituting Eq.~\eqref{eq:int 1} into the definition of $S_{1}$ in Eq.~\eqref{eq:def of lat sum 2d}, we obtain
\begin{equation}
    S_{1+}(\alpha,\bm{\beta}) = \frac{1}{ 2 \pi^{2}} \int_{0}^{1} dt ~ t^{\bm{\delta}^{2} - 1} \sum_{ \mathbf{R} } t^{\mathbf{R}^{2} + 2 \mathbf{R} \cdot \bm{\delta}} {e}^{- 2 \mathrm{i} \pi \bm{\beta} \cdot \mathbf{R}}.
    \label{eq: int1 of S1}
\end{equation} 
We now write the summation explicitly:
\begin{align}
    &\sum_{\mathbf{R}}t^{\mathbf{R}^{2} + 2 \mathbf{R} \cdot \bm{\delta}} {e}^{- 2 \mathrm{i} \pi \bm{\beta} \cdot \mathbf{R}} \nonumber\\
    = &\sum\limits_{(j_{1},j_{2}) \in \mathbb{Z}^{2}} t^{\frac{3}{4} (j_{1}-j_{2})^{2}  + \frac{1}{4}(j_{1}+j_{2})^{2} + \sqrt{3} \delta_{x} (j_{1}-j_{2}) + \delta_{y}(j_{1}+j_{2})} {e}^{- \mathrm{i} \pi \left[ \sqrt{3} (j_{1}-j_{2}) \beta_{x} + (j_{1}+j_{2}) \beta_{y} \right]},
    \label{eq: sum of theta}
\end{align}
Here we define the variables $\delta_{x} = \sqrt{3} (\delta_{1} - \delta_{2}) / 2$, $\delta_{y} = - (\delta_{1} + \delta_{2}) / 2$, and $\beta_{x} = (\beta_{1} + \beta_{2}) / \sqrt{3}$, $\beta_{y} = \beta_{1} - \beta_{2}$. 

We observe that $j_{1} - j_{2}$ and $j_{1} + j_{2}$ share the same parity (they are either both even or both odd). We thus divide the lattice sum into two contributions. For the even case, we make the substitution $(j_{1} - j_{2}, j_{1} + j_{2}) \to (2n_{1}, 2n_{2})$; for the odd case, we use $(j_{1} - j_{2}, j_{1} + j_{2}) \to (2n_{1}+1, 2n_{2}+1)$. Therefore, we obtain
\begin{align}
    &\sum_{\mathbf{R}}t^{\mathbf{R}^{2} + 2 \mathbf{R} \cdot \bm{\delta}} {e}^{- 2 \mathrm{i} \pi \bm{\beta} \cdot \mathbf{R}}  = \sum\limits_{n_{1},n_{2}} t^{ \mathbf{n}^{2} + 2 \mathbf{n} \cdot \bm{\delta}} {e}^{-2 \mathrm{i} \pi  \bm{\beta} \cdot \mathbf{n} } + (n_{1},n_{2}) \to  (n_{1} + 1 / 2, n_{2} +  1 / 2),
    \label{eq: sum of theta 2}
\end{align}
where we defined $\mathbf{n} = (\sqrt{3} n_{1},n_{2})$ with $n_{1}, n_{2} \in \mathbb{Z}$. 

We note that the second and third kind of Jacobi theta functions are defined as $\vartheta_{2}(z,t) = \sum_{n= -\infty}^{\infty} t^{(n+\frac{1}{2})^{2}} {e}^{(2 n+1) \mathrm{i} z } $ and $\vartheta_{3}(z,t) = \sum_{n= -\infty}^{\infty} t^{n^{2}} {e}^{ 2 n \mathrm{i} z } $. Thus, the right-hand side of Eq.~\eqref{eq: sum of theta 2} can be expressed in terms of Jacobi theta functions as
\begin{align}
     &\sum\limits_{\mathbf{n}} t^{ \mathbf{n}^{2} + 2 \mathbf{n} \cdot \bm{\delta}} {e}^{-2 \mathrm{i} \pi  \bm{\beta} \cdot \mathbf{n} } + (n_{1},n_{2}) \to  (n_{1} + 1 / 2, n_{2} +  1 / 2) \nonumber \\
    &= \vartheta_{3}( \sqrt{3} ( \pi \beta_{x} + \mathrm{i} \delta_{x} \ln  t),  t^{3}) \vartheta_{3}(\pi \beta_{y} + \mathrm{i}\delta_{y} \ln t,  t) + (\vartheta_{3} \to  \vartheta_{2}).
    \label{eq: formula of lattice sum}
\end{align}
Substituting Eq.~\eqref{eq: formula of lattice sum} into Eq.~\eqref{eq: int1 of S1}, we arrive at:
\begin{equation}
    S_{1+}(\alpha,\bm{\beta}) = \frac{1}{ 2 \pi^{2}} \int_{0}^{1} dt ~ t^{\bm{\delta}^{2} - 1} \left[ \vartheta_{3}( \sqrt{3} ( \pi \beta_{x} + \mathrm{i} \delta_{x} \ln  t),  t^{3}) \vartheta_{3}(\pi \beta_{y} + \mathrm{i}\delta_{y} \ln t,  t) + (\vartheta_{3} \to  \vartheta_{2})  \right].
    \label{eq: s1 2d}
\end{equation}

\subsection{$S_{2+}$ in 2D}
The calculation of $S_{2+}(\alpha,\bm{\beta})$ parallels that of $S_{1-}(\alpha,\bm{\beta})$, but requires a different auxiliary integral to express the summand in terms of theta functions. This integral is given by
\begin{equation}
    - \frac{1}{2\sqrt{\pi}} \int_{0}^{1} dt ~ \frac{ {e}^{- \frac{\pi^{2} \alpha^{2}}{\ln t }}}{ \sqrt{ - \ln t }} \text{erfc}(\frac{\pi \alpha}{ \sqrt{ -  \ln t }}) t^{ x^{2} - 1} = \frac{1}{2\pi \mathrm{i} x} \left[ {e}^{2 \mathrm{i} \pi \alpha x} {e}_{1} (2 \mathrm{i} \pi \alpha x) - (\alpha \to  -\alpha)\right].
\end{equation}
This identity can be verified by applying the Mellin transform with respect to the variable $\alpha$ on the right-hand side. Following the same steps as in the evaluation of $S_{1+}(\alpha,\bm{\beta})$, we obtain the expression for $S_{2+}(\alpha,\bm{\beta})$:
\begin{align}
    S_{2+}(\alpha,\bm{\beta}) = &- \frac{\alpha}{2 \sqrt{\pi}} \int_{0}^{1} dt ~ \frac{{e}^{ -  \frac{ \pi^{2} \alpha^{2}}{ \ln t }}}{ \sqrt{ - \ln t}} \text{erfc}( \frac{\pi \alpha}{ \sqrt{ -  \ln t}} )t^{\bm{\delta}^{2} - 1} \times \nonumber \\
    &\left[ \vartheta_{3}(\sqrt{3} (\pi \beta_{x}  + \mathrm{i} \delta_{x} \ln t),  t^{3}) \vartheta_{3}(\pi \beta_{y} + \mathrm{i}\delta_{y} \ln t,  t) + (\vartheta_{3} \to  \vartheta_{2})  \right].
    \label{eq: s2 2d}
\end{align}

\subsection{$S_{3+}$ in 2D}
We note that the final term admits the following Ewald integral representation \cite{beylkin2008fast}:
\begin{equation}
    \frac{2}{\sqrt{\pi}} \int_{(0)}^{\infty} \frac{du}{u^{2}} {e}^{\pi^{2} \alpha^{2} u^{2}  - x^{2} / u^{2}} = \frac{{e}^{2 \mathrm{i} \pi \alpha x }}{ x },
    \label{eq: Ewald rep}
\end{equation}
Here the integral is taken along a contour in the complex plane chosen to ensure convergence \cite{capolino2007efficient}. Consequently, $S_{3+}(\alpha,\bm{\beta})$ can be expressed as
\begin{align}
    S_{3+}(\alpha,\bm{\beta}) = &\frac{2 \alpha}{\sqrt{\pi}} \int_{(0)}^{\infty} \frac{du}{u^{2}} {e}^{ \pi^{2} \alpha^{2} u^{2}  - \bm{\delta}^{2} / u^{2}}  \nonumber\\
    &\times \left[ \vartheta_{3}( \sqrt{3} (\pi \beta_{x}- \mathrm{i} \delta_{x} / u^{2}), {e}^{- 3 / u^{2}}) \vartheta_{3}(\pi \beta_{y}- \mathrm{i} \delta_{y} / u^{2}, {e}^{- 1 / u^{2}})  + (\vartheta_{3} \to  \vartheta_{2}) \right].
    \label{eq: s3 2d}
\end{align}

\subsection{Result Expressed in Terms of Theta Functions}
Collecting the results given in Eqs. \eqref{eq: s1 2d}, \eqref{eq: s2 2d} and \eqref{eq: s3 2d}, we obtain the final result for $S_{+}$:
\begin{align}
    &S_{+}(\alpha,\bm{\beta}) =  \nonumber\\
    &\int_{0}^{1} dt \left( \frac{1}{2 \pi^{2} } - \frac{\alpha {e}^{- \frac{\pi^{2} \alpha^{2} }{\ln t }}  }{2  \sqrt{ - \pi \ln t } } \text{erfc}(\frac{ \pi \alpha}{ \sqrt{- \ln t} })  \right) t^{ \bm{\delta}^{2} - 1} \vartheta_{2}( \sqrt{3} (\pi \beta_{x} + \mathrm{i} \delta_{x} \ln t),t^{3}) \vartheta_{2}(\pi \beta_{y} +\mathrm{i} \delta_{y} \ln t,t)  \nonumber\\
    & + \frac{2 \alpha}{\sqrt{\pi}}\int_{(0)}^{\infty} \frac{du}{u^{2}} {e}^{\pi^{2} \alpha^{2} u^{2} - \bm{\delta}^{2} / u^{2} } \vartheta_{2}( \sqrt{3} (\pi \beta_{x} - \mathrm{i} \delta_{x} / u^{2}),{e}^{-\frac{3}{u^{2}}}) \vartheta_{2}(\pi \beta_{y} - \mathrm{i} \delta_{y} / u^{2},{e}^{-\frac{1}{u^{2}}}) + ( \vartheta_{2} \to \vartheta_{3} )  .
    \label{eq:s+ 2d}
\end{align}
For the hexagonal lattice, we set $\delta_{x} = -1 / \sqrt{3}$ and $\delta_{y} = 0$. 

The expression for $S_{-}$ is obtained by replacing $\bm{\delta}$ with $-\bm{\delta}$, while the expression of $S_{0}$ is derived by subtracting the divergent $\mathbf{R} = 0$ term from $S_{+}$ and taking the limit $\bm{\delta} \to 0$, resulting in
\begin{align}
    &S_{0}(\alpha,\bm{\beta}) =  \nonumber\\
    &\int_{0}^{1} dt \left( \frac{1}{2 \pi^{2} t } - \frac{\alpha {e}^{- \frac{\pi^{2} \alpha^{2} }{\ln t }} }{2 t \sqrt{ - \pi \ln t } } \text{erfc}(\frac{ \pi \alpha}{ \sqrt{- \ln t} })  \right) [ \vartheta_{2}(\sqrt{3} \pi \beta_{x} ,t^{3}) \vartheta_{2}(\pi \beta_{y},t)  - \frac{1}{2}]  \nonumber\\
    & + \frac{2 \alpha}{\sqrt{\pi}}\int_{(0)}^{\infty} \frac{du}{u^{2}} {e}^{\pi^{2} \alpha^{2} u^{2} } [ \vartheta_{2}( \sqrt{3} \pi \beta_{x},{e}^{-\frac{3}{u^{2}}}) \vartheta_{2}(\pi \beta_{y},{e}^{-\frac{1}{u^{2}}}) -\frac{1}{2}] + (\vartheta_{2} \to  \vartheta_{3}) .
    \label{eq:s0 2d}
\end{align}
The first term in the integrand of Eqs.~\eqref{eq:s+ 2d} and \eqref{eq:s0 2d} converges rapidly and accurately. In contrast, the second term exhibits strong oscillations near $u=0$ and $u=\infty$, posing challenges for numerical evaluation. In the following section, we present a derivation of a numerically efficient and stable formula for the last term in the expressions of $S_{+}$ and $S_{0}$.

\subsection{\label{appd:ewald sum}Ewald Summation Method}
Here we derive a numerically efficient and stable expression for the lattice sums using the Ewald summation method \cite{linton1998green,beylkin2008fast,beutel2023unified}. We begin by writing
\begin{equation}
    \frac{2}{\sqrt{\pi}} \int_{(0)}^{\infty} \frac{du}{u^{2}} {e}^{\pi^{2} \alpha^{2} u^{2}  - x^{2} / u^{2}}  = \frac{2}{\sqrt{\pi}} \left( \int_{(0)}^{\eta} + \int_{\eta}^{\infty} \right)  \frac{du}{u^{2}} {e}^{\pi^{2} \alpha^{2} u^{2}  - x^{2} / u^{2}}.
\end{equation}
The key idea is to separate the integration range over $u$ into two parts: one covering small $u$ values and the other covering large $u$. The former converges rapidly in real space, while the latter converges quickly in Fourier space. Accordingly, the lattice sum $S_{3+}$ is divided as 
\begin{subequations}
    \begin{align}
        S_{3+}^{r}(\alpha,\bm{\beta}) &= \frac{2 \alpha}{ \sqrt{\pi} } \sum_{\mathbf{R}} \int_{(0)}^{\eta} \frac{du}{u^{2}} {e}^{\pi^{2} \alpha^{2} u^{2} - \left\vert \mathbf{R} + \bm{\delta} \right\vert ^{2} / u^{2}} {e}^{-2 \mathrm{i} \pi \bm{\beta} \cdot \mathbf{R}}, \\
        S_{3+}^{m}(\alpha,\bm{\beta}) &= \frac{2 \alpha}{ \sqrt{\pi} } \sum_{\mathbf{R}} \int_{\eta}^{\infty} \frac{du}{u^{2}} {e}^{\pi^{2} \alpha^{2} u^{2} - \left\vert \mathbf{R} + \bm{\delta} \right\vert ^{2} / u^{2}} {e}^{-2 \mathrm{i} \pi \bm{\beta} \cdot \mathbf{R}},
    \end{align}
    \label{eq:ewald expression}
\end{subequations}
where $\eta$ is a parameter.

For $S_{3+}^{r}$, the integral can be evaluated explicitly:
\begin{equation}
    \int_{0}^{\eta} \frac{du}{u^{2}} {e}^{\pi^{2} \alpha^{2} u^{2} - x^{2} / u^{2}} = \frac{\sqrt{\pi}}{4 x} \left[ {e}^{2 \mathrm{i} \pi \alpha x} \text{erfc}(\frac{x}{\eta} + \mathrm{i} \pi \alpha \eta) + {e}^{-2 \mathrm{i} \pi \alpha x} \text{erfc}(\frac{x}{\eta} - \mathrm{i} \pi \alpha \eta) \right].
    \label{eq:int real}
\end{equation}
Substituting Eq.~\eqref{eq:int real} into Eq.~\eqref{eq:ewald expression}, we can express $S_{3+}^{r}$ as
\begin{align}
    S_{3+}^{r}(\alpha,\bm{\beta}) =\frac{\alpha}{2} \sum_{\mathbf{R}} \frac{{e}^{- 2 \mathrm{i} \pi \bm{\beta} \cdot \mathbf{R}}}{ \left\vert  \mathbf{R} + \bm{\delta} \right\vert } \left[ {e}^{2 \mathrm{i} \pi \alpha \left\vert  \mathbf{R} + \bm{\delta} \right\vert} \text{erfc}(\frac{\left\vert  \mathbf{R} + \bm{\delta} \right\vert}{\eta} + \mathrm{i} \pi \alpha \eta) + (\alpha \to  - \alpha)\right].
    \label{eq:s3+ real}
\end{align}
On the other hand, the sum $S_{3+}^{m}(\alpha,\bm{\beta})$ converges rapidly in Fourier space. Applying the Poisson summation formula we obtain
\begin{equation}
    S_{3+}^{m}(\alpha,\bm{\beta}) = 16 \alpha \sqrt{\frac{\pi}{3} } \sum_{\mathbf{G}} \int_{\eta}^{\infty} du ~{e}^{\pi^{2} u^{2} \left( \alpha^{2} - \left\vert \bm{\beta} + \mathbf{G} \right\vert^{2}  \right) + 2 \mathrm{i} \pi \bm{\delta}\cdot \left( \bm{\beta} + \mathbf{G} \right)}.
    \label{eq:s3+ reci 0}
\end{equation}
Here $\mathbf{G} = m \mathbf{b}_{1} + n \mathbf{b}_{2}$, where $m,n \in \mathbb{Z}$ denote the reciprocal lattice vectors of the honeycomb lattice. The integral in Eq.~\eqref{eq:s3+ reci 0} is related to the complementary error function, leading to
\begin{equation}
    S_{3+}^{m}(\alpha,\bm{\beta})  = \frac{8 \alpha}{\sqrt{3}} \sum_{\mathbf{G}}\frac{\text{erfc}\left( \pi \eta \sqrt{ \left\vert \bm{\beta} + \mathbf{G} \right\vert^{2} - \alpha^{2}  } \right)}{ \sqrt{ \left\vert \bm{\beta} + \mathbf{G} \right\vert^{2} - \alpha^{2}  } } {e}^{2 \mathrm{i} \pi \bm{\delta} \cdot \left( \bm{\beta} + \mathbf{G} \right)}.
    \label{eq:s3+ reci}
\end{equation}
Combining Eqs.~\eqref{eq:s3+ real} and \eqref{eq:s3+ reci} gives the Ewald summation formula, which, when substituted into Eq.~\eqref{eq:s+ 2d} yields
\begin{align}
    &S_{+}(\alpha,\bm{\beta}) = \nonumber \\ 
    &\int_{0}^{1} dt \left( \frac{1}{2 \pi^{2} } - \frac{\alpha {e}^{- \frac{\pi^{2} \alpha^{2} }{\ln t }}  }{2  \sqrt{ - \pi \ln t } } \text{erfc}(\frac{ \pi \alpha}{ \sqrt{- \ln t} })  \right) t^{ \bm{\delta}^{2} - 1}  \vartheta_{2}(\widetilde{\beta}_{x}(t) ,t^{3}) \vartheta_{2}(\widetilde{\beta}_{y}(t),t)  + (\vartheta_{2} \leftrightarrow \vartheta_{3}) \nonumber\\
    &+\frac{\alpha}{2} \sum_{\mathbf{R}} \frac{{e}^{- 2 \mathrm{i} \pi \bm{\beta} \cdot \mathbf{R}}}{ \left\vert  \mathbf{R} + \bm{\delta} \right\vert } \left[ {e}^{2 \mathrm{i} \pi \alpha \left\vert  \mathbf{R} + \bm{\delta} \right\vert} \text{erfc}(\frac{\left\vert  \mathbf{R} + \bm{\delta} \right\vert}{\eta} + \mathrm{i} \pi \alpha \eta) + (\alpha \to  - \alpha)\right] \nonumber \\
    &+ \frac{8 \alpha}{\sqrt{3}} \sum_{\mathbf{G}}\frac{\text{erfc}\left( \pi \eta \sqrt{ \left\vert \bm{\beta} + \mathbf{G} \right\vert^{2} - \alpha^{2}  } \right)}{ \sqrt{ \left\vert \bm{\beta} + \mathbf{G} \right\vert^{2} - \alpha^{2}  } } {e}^{2 \mathrm{i} \pi \bm{\delta} \cdot \left( \bm{\beta} + \mathbf{G} \right)},
    \label{eq:s+ ewald}
\end{align}
where $\widetilde{\beta}_{x}(t) = \sqrt{3} (\pi \beta_{x} + \mathrm{i} \delta_{x} \ln t)$ and $\widetilde{\beta}_{y}(t) = \pi \beta_{y} + \mathrm{i} \delta_{y} \ln t$. Both sums in Eq.~\eqref{eq:s+ ewald} converge exponentially, so only a few terms are needed for an accurate evaluation of the lattice sum.

The Ewald summation expression for $S_{0}(\alpha,\bm{\beta})$ is obtained by excluding the divergent $\mathbf{R} = 0$ term from the real-space sum in Eq.~\eqref{eq:s+ ewald} and taking the limit $\bm{\delta} \to 0$, resulting in
\begin{align}
     &S_{0}(\alpha,\bm{\beta}) =  \nonumber\\
    &\int_{0}^{1} dt \left( \frac{1}{2 \pi^{2} t } - \frac{\alpha {e}^{- \frac{\pi^{2} \alpha^{2} }{\ln t }} }{2 t \sqrt{ - \pi \ln t } } \text{erfc}(\frac{ \pi \alpha}{ \sqrt{- \ln t} })  \right) [ \vartheta_{2}(\sqrt{3} \pi \beta_{x} ,t^{3}) \vartheta_{2}(\pi \beta_{y},t)  - \frac{1}{2}] +(\vartheta_{2} \leftrightarrow \vartheta_{3}) \nonumber\\
     &+\frac{\alpha}{2} \sum_{\mathbf{R} \neq 0} \frac{{e}^{- 2 \mathrm{i} \pi \bm{\beta} \cdot \mathbf{R}}}{ \left\vert  \mathbf{R} \right\vert } \left[ {e}^{2 \mathrm{i} \pi \alpha \left\vert  \mathbf{R} \right\vert} \text{erfc}(\frac{\left\vert  \mathbf{R} \right\vert}{\eta} + \mathrm{i} \pi \alpha \eta) +   {e}^{-2 \mathrm{i} \pi \alpha \left\vert  \mathbf{R} \right\vert} \text{erfc}(\frac{\left\vert  \mathbf{R} \right\vert}{\eta} - \mathrm{i} \pi \alpha \eta) \right] \nonumber \\
    &+ \frac{8 \alpha}{\sqrt{3}} \sum_{\mathbf{G}}\frac{\text{erfc}\left( \pi \eta \sqrt{ \left\vert \bm{\beta} + \mathbf{G} \right\vert^{2} - \alpha^{2}  } \right)}{ \sqrt{ \left\vert \bm{\beta} + \mathbf{G} \right\vert^{2} - \alpha^{2}  } } - \frac{2 \alpha}{\sqrt{\pi} \eta} {e}^{\pi^{2} \alpha^{2} \eta^{2}}- 2 \mathrm{i} \pi \alpha^{2} \text{erfc}(-\mathrm{i} \pi \alpha \eta).
    \label{eq:s0 ewald}
\end{align}
Although the parameter $\eta$ is an arbitrary real constant, it is typically chosen to balance the convergence rates of the real- and reciprocal-space sums. We have numerically evaluated Eqs.~\eqref{eq:s+ ewald} and \eqref{eq:s0 ewald} for various values of $\eta$ and have confirmed that, as expected, the results are independent of this choice.

\section{\label{appd:symmetry of lat sum}Formation of Dirac Cones}
Here we derive the symmetry properties of the lattice sums that are essential for establishing the low-energy Dirac Hamiltonian discussed in the main text. We demonstrate that the formation of Dirac cones is a direct consequence of lattice symmetries.

\subsection{Quasi-Inversion Symmetry of the Lattice Sums}
\begin{figure}
    \centering
    \includegraphics[width=0.8\textwidth]{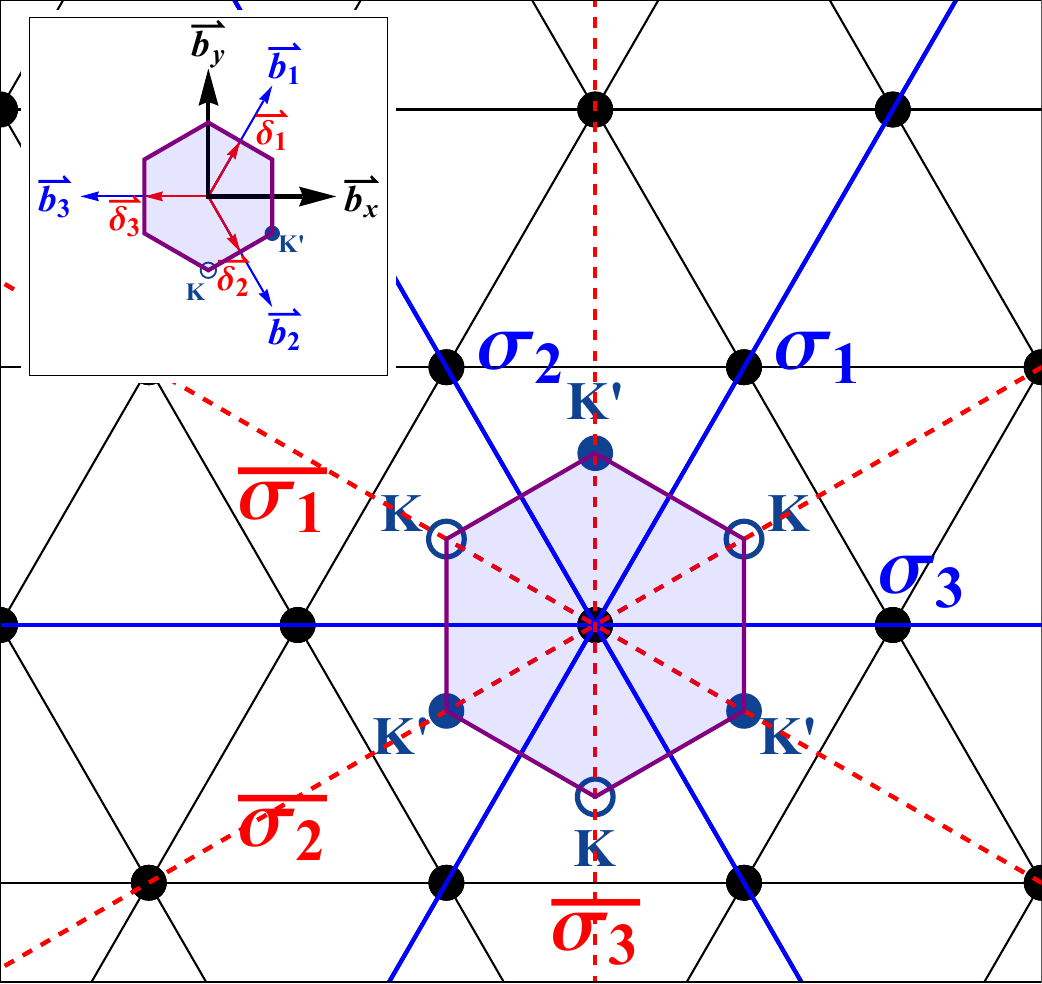}
    \caption{Reciprocal lattice of the honeycomb lattice. The black dots indicate the reciprocal lattice points, and the shaded hexagon represents the FBZ. The high-symmetry Dirac points, $K$ (open circle) and $K^{\prime}$ (closed circle), are indicated. The inset shows the reciprocal lattice basis vectors ($\mathbf{b}_{1},~\mathbf{b}_{2},~\mathbf{b}_{3}$), the standard Cartesian basis ($\mathbf{b}_{x},~ \mathbf{b}_{y}$), and the real-space displacement vectors ($\bm{\delta}_i$, red) connecting the two sublattices. Also shown are the mirror planes $\sigma_{i}$ and $\bar{\sigma}_{i}$, which are parallel and perpendicular to the basis vectors $\mathbf{b}_{i}$, respectively.}
    \label{fig:reciprocal lattice}
\end{figure}

First, we demonstrate the quasi-inversion symmetry of the lattice sum $S_{\pm}(\bm{\beta})$ around the Dirac points $K$ and $K^{\prime}$. We begin with the definition of the lattice sum, suppressing the energy dependence of the Green's function for notational simplicity:
\begin{equation} 
    S_{\pm}(\bm{\beta}) = \sum_{\mathbf{R}} G(|\mathbf{R} \pm \bm{\delta}|) {e}^{- 2 \mathrm{i} \pi \bm{\beta} \cdot \mathbf{R}}. 
    \label{eq:lattice sum appd} 
\end{equation}
Here $\mathbf{R} = m \mathbf{a}_{1}+ n \mathbf{a}_{2}$ (with $m,n \in \mathbb{Z}$) are the real-space lattice vectors. We note the following symmetry relation:
\begin{equation} 
    S_{\pm}(\bm{\beta}_{K (K^{\prime})}+ q \mathbf{b}_{i}) = S_{\mp}(\bm{\beta}_{K(K^{\prime})} - q \mathbf{b}_{i}) {e}^{\pm 2\pi \mathrm{i} \bm{\beta}_{K(K^{\prime})} \cdot ( \bm{\delta} - \hat{\sigma}_{i} \bm{\delta})}. 
    \label{eq:quasi-inversion symmetry} 
\end{equation}
The terms in this expression are defined with respect to the reciprocal lattice shown in Fig.~\ref{fig:reciprocal lattice}. The vectors $\bm{\beta}_{K}$ and $\bm{\beta}_{K^{\prime}}$ denote the positions of the Dirac points. The operator $\hat{\sigma}_{i}$ represents a mirror reflection across a plane parallel to the reciprocal basis vector $\mathbf{b}_{i}$. This operation is a symmetry of the lattice that exchanges the two inequivalent Dirac points, $\hat{\sigma}_{i} \bm{\beta}_{K} = \bm{\beta}_{K^{\prime}}$, which are also related by inversion, $\bm{\beta}_{K} = - \bm{\beta}_{K^{\prime}}$.

Using the inversion symmetry of the lattice sum, $S_{\pm}(\bm{\beta}) = S_{\mp}(- \bm{\beta})$, and the relation $\bm{\beta}_{K^{\prime}} = -\bm{\beta}_{K}$, the symmetry shown in Eq.~\eqref{eq:quasi-inversion symmetry} can be rewritten in an equivalent form:
\begin{equation} 
    S_{\pm}(\bm{\beta}_{K}+ q \mathbf{b}_{i}) = S_{\pm}(\bm{\beta}_{K^{\prime}} + q \mathbf{b}_{i}) {e}^{\pm 2\pi \mathrm{i} \bm{\beta}_{K} \cdot ( \bm{\delta} - \hat{\sigma}_{i} \bm{\delta})}. 
    \label{eq:quasi-inversion symmetry 2} 
\end{equation}
We therefore focus on demonstrating this equivalence.
The derivation proceeds by transforming the lattice sum in Eq.~\eqref{eq:lattice sum appd} into Fourier space using the Poisson summation formula:
\begin{equation} 
    S_{\pm}(\bm{\beta}) = C \sum_{\mathbf{Q} \in \Lambda^*} \widetilde{G}(|\mathbf{Q} +\bm{\beta}|) {e}^{\pm 2 \pi \mathrm{i}  (\mathbf{Q}+ \bm{\beta}) \cdot \bm{\delta} }. 
    \label{eq:reciprocal space} 
\end{equation}
Here $C$ is a constant related to the volume of the Brillouin zone, $\Lambda^*$ is the reciprocal lattice, and $\widetilde{G}$ denotes the Fourier transform of $G$. Evaluating this expression at a point $\bm{\beta} = \bm{\beta}_{K^{\prime}} + q \mathbf{b}_{i}$ near $K^{\prime}$ gives
\begin{equation} 
    S_{\pm}(\bm{\beta}_{K^{\prime}} + q \mathbf{b}_{i}) = C \sum_{\mathbf{Q} \in \Lambda^*} \widetilde{G}(|\mathbf{Q} +\bm{\beta}_{K^{\prime}} +q \mathbf{b}_{i}|) {e}^{\pm 2 \pi \mathrm{i}  (\mathbf{Q}+ \bm{\beta}_{K^{\prime}} +q \mathbf{b}_{i}) \cdot \bm{\delta} }. 
    \label{eq:reciprocal space K'} 
\end{equation}
We now apply the mirror symmetry $\hat{\sigma}_{i}$. Since this operator preserves the inner product ($\mathbf{x}\cdot \mathbf{y}  = (\hat{\sigma}_{i} \mathbf{x}) \cdot (\hat{\sigma}_{i} \mathbf{y})$), we can transform the terms inside the sum:
\begin{equation} 
    S_{\pm}(\bm{\beta}_{K^{\prime}} + q \mathbf{b}_{i})  = C \sum_{\mathbf{Q} \in \Lambda^*} \widetilde{G}(| \hat{\sigma}_{i} (\mathbf{Q} +\bm{\beta}_{K^{\prime}} +q \mathbf{b}_{i} ) |) {e}^{\pm 2 \pi \mathrm{i}  [\hat{\sigma}_{i} (\mathbf{Q}+ \bm{\beta}_{K^{\prime}} +q \mathbf{b}_{i})]  \cdot  (\hat{\sigma}_{i} \bm{\delta})  }. 
\end{equation}
The operator $\hat{\sigma}_{i}$ acts on  vectors as follows: it maps the reciprocal lattice onto itself ($\hat{\sigma}_{i} \mathbf{Q} = \mathbf{Q}^{\prime} \in \Lambda^*$), it exchanges the Dirac points ($\hat{\sigma}_{i} \bm{\beta}_{K^{\prime}} = \bm{\beta}_{K}$), and it leaves the parallel basis vector unchanged ($\hat{\sigma}_{i} \mathbf{b}_{i} = \mathbf{b}_{i}$). Applying these properties yields
\begin{equation} 
    S_{\pm}(\bm{\beta}_{K^{\prime}} + q \mathbf{b}_{i}) = C \sum_{\mathbf{Q}^{\prime} \in \Lambda^*} \widetilde{G}(|\mathbf{Q}^{\prime} +\bm{\beta}_{K} +q \mathbf{b}_{i}|) {e}^{\pm 2 \pi \mathrm{i}  (\mathbf{Q}^{\prime}+ \bm{\beta}_{K} +q \mathbf{b}_{i}) \cdot \hat{\sigma}_{i} \bm{\delta} }. 
    \label{eq:reciprocal space K' 2} 
\end{equation}
In addition, the vector $\bm{\delta}$ connects an A-sublattice site to a B-sublattice site. The mirror operation, being a symmetry operation of the lattice, must map this B-site to another B-site. This can be expressed as $\hat{\sigma}_{i} \bm{\delta} = \bm{\delta} + \mathbf{R}$ for some real-space lattice vector $\mathbf{R} \in \Lambda$. This implies a crucial simplification for the phase factor involving $\mathbf{Q}^{\prime}$:
\begin{equation} 
    {e}^{ 2\pi \mathrm{i} \mathbf{Q}^{\prime} \cdot \hat{\sigma}_{i} \bm{\delta} } = {e}^{2 \pi \mathrm{i} (\mathbf{Q}^{\prime} \cdot \bm{\delta} + \mathbf{Q}^{\prime} \cdot \mathbf{R} ) } = {e}^{2 \pi \mathrm{i} \mathbf{Q}^{\prime} \cdot \bm{\delta} }, 
    \label{eq:lemma 1} 
\end{equation}
because $\mathbf{Q}^{\prime} \cdot \mathbf{R}$ is always an integer. Substituting Eq.~\eqref{eq:lemma 1} into Eq.~\eqref{eq:reciprocal space K' 2} and rearranging the remaining phase factors, we find:
\begin{align} 
    S_{\pm}(\bm{\beta}_{K^{\prime}} + q \mathbf{b}_{i}) &= \left[c \sum_{\mathbf{Q}^{\prime} \in \Lambda^*} \widetilde{G}(|\mathbf{Q}^{\prime} +\bm{\beta}_{K} +q \mathbf{b}_{i}|) {e}^{\pm 2 \pi \mathrm{i}   (\mathbf{Q}^{\prime} + \bm{\beta}_{K} + q \mathbf{b}_{i}) \cdot  \bm{\delta}} \right] {e}^{ \mp 2 \pi \mathrm{i} ( \bm{\beta}_{K} \cdot \bm{\delta} -  \bm{\beta}_{K} \cdot (\hat{\sigma}_{i}\bm{\delta})) }, \nonumber \\
     & = S_{\pm}(\bm{\beta}_{K} + q \mathbf{b}_{i}) {e}^{ \mp 2 \pi \mathrm{i} ( \bm{\beta}_{K} - \hat{\sigma}_{i} \bm{\beta}_{K})\cdot \bm{\delta} }. 
\end{align}
This result is precisely the relation in Eq.~\eqref{eq:quasi-inversion symmetry 2}.

For the specific choice of basis vectors used in the main text and shown in Fig.~\ref{fig:reciprocal lattice}, the general symmetry relation in Eq.~\eqref{eq:quasi-inversion symmetry} takes the explicit form:
\begin{subequations} 
    \begin{align} S_{+}(\bm{\beta}_{K} + q \mathbf{b}_{1}) &= S_{-}(\bm{\beta}_{K} - q \mathbf{b}_{1}) {e}^{-  \mathrm{i} \frac{2}{3} \pi }, \\ S_{+}(\bm{\beta}_{K} + q \mathbf{b}_{2}) &= S_{-}(\bm{\beta}_{K} - q \mathbf{b}_{2}) {e}^{  \mathrm{i} \frac{2}{3} \pi }, \\ S_{+}(\bm{\beta}_{K} + q \mathbf{b}_{3}) &= S_{-}(\bm{\beta}_{K} - q \mathbf{b}_{3}). \end{align}  
    \label{eqs:quasi-inversion symmetry 3} 
\end{subequations}
These relations can be readily verified using Eq.~\eqref{eq:s+ ewald}.

We conclude by emphasizing the generality of the result. The proof relies only on the lattice symmetries and does not assume isotropy—that is, it does not require the Green's function take the rotationally invariant form $G(|\mathbf{r}|)$. Consequently, the quasi-inversion symmetry remains valid even in the presence of anisotropic interactions, provided these interactions respect the point group symmetry of the honeycomb lattice.

\subsection{Band Gap is Closed at Dirac Points}
A direct consequence of the quasi-inversion symmetry in Eq.~\eqref{eqs:quasi-inversion symmetry 3} is that the off-diagonal components of the Hamiltonian vanish at the Dirac points. The value of $S_{+}(\bm{\beta}_{K})$ can be determined by averaging over the three symmetric directions in the limit $q \to 0$:
\begin{align} 
    S_{+}(\bm{\beta}_{K}) &= \frac{1}{3} \lim_{q \to 0}  \left[ S_{+}(\bm{\beta}_{K} + q \mathbf{b}_{1}) +  S_{+}(\bm{\beta}_{K} + q \mathbf{b}_{2}) +  S_{+}(\bm{\beta}_{K} + q \mathbf{b}_{3}) \right]  \nonumber \\ 
    & = \frac{1}{3} S_{-}(\bm{\beta}_{K})\left({e}^{-  \mathrm{i} \frac{2}{3} \pi } + {e}^{  \mathrm{i} \frac{2}{3} \pi } + 1\right) = 0. 
\end{align}
An identical argument shows that $S_{-}(\bm{\beta}_{K})=0$. From the definitions in Eq.~\eqref{eq:def of h}, it immediately follows that $h_{1}(\bm{\beta}_{K(K^{\prime})}) = h_{2}(\bm{\beta}_{K(K^{\prime})}) = 0$. Consequently, for a system with preserved chiral symmetry (i.e., when $h_{3}=0$), the two energy bands become degenerate at the Dirac points, $\alpha_{\pm}(\bm{\beta}_{K(K^{\prime})}) = h_{0}(\bm{\beta}_{K(K^{\prime})})$, confirming that the band gap is necessarily closed.

\subsection{\label{appd:low energy expansion}Fermi Velocity is Isotropic}
Finally, as an application of the quasi-inversion symmetry, we show that the Fermi velocity is isotropic and derive the low-energy expansion of the Hamiltonian around the Dirac points. We expand the momentum as $\bm{\beta} = \bm{\beta}_{K} +\mathbf{q}$, where $\mathbf{q} = q_{x} \mathbf{b}_{x} + q_{y} \mathbf{b}_{y}$ is the small deviation from the $K$ point in the orthonormal Cartesian basis. Since $h_{1}(\bm{\beta}_{K})= h_{2}(\bm{\beta}_{K}) = 0$, the expansion begins at linear order:
\begin{subequations} 
    \begin{align} h_{1}(\bm{\beta}_{K} +\mathbf{q}) &=  \frac{\partial h_{1}}{\partial \beta_{x}} \bigg\rvert_{\bm{\beta}= \bm{\beta}_{K}} q_{x}+ \frac{\partial h_{1}}{\partial \beta_{y}} \bigg\rvert_{\bm{\beta}= \bm{\beta}_{K}} q_{y} + \mathcal{O}(\mathbf{q}^2), \\ 
    h_{2}(\bm{\beta}_{K} +\mathbf{q}) &=  \frac{\partial h_{2}}{\partial \beta_{x}} \bigg\rvert_{\bm{\beta}= \bm{\beta}_{K}} q_{x}+ \frac{\partial h_{2}}{\partial \beta_{y}} \bigg\rvert_{\bm{\beta}= \bm{\beta}_{K}} q_{y} + \mathcal{O}(\mathbf{q}^2). 
    \end{align} 
\end{subequations}
To leverage the system's symmetries, which are expressed naturally in the reciprocal lattice basis $\{\mathbf{b}_{1},\mathbf{b}_{2}\}$, we relate the Cartesian derivatives to the derivatives with respect to $\beta_1$ and $\beta_2$ via a change of basis:
\begin{subequations} 
    \begin{align} \frac{\partial h_{1}}{\partial \beta_{x}}  &=  - \frac{\sqrt{3}}{4} \sqrt{\kappa_{A} \kappa_{B}}  \left[ \frac{\partial  }{\partial \beta_{1}} (S_{+}+ S_{-}) + \frac{\partial  }{\partial \beta_{2}} (S_{+}+ S_{-}) \right] , \\ \frac{\partial h_{1}}{\partial \beta_{y}}  
    &=  - \frac{1}{4} \sqrt{\kappa_{A} \kappa_{B}}  \left[ \frac{\partial  }{\partial \beta_{1}} (S_{+}+ S_{-}) - \frac{\partial  }{\partial \beta_{2}} (S_{+}+ S_{-}) \right] , \label{eq:derivative 2} \\ 
    \frac{\partial h_{2}}{\partial \beta_{x}}  &=  \frac{\sqrt{3}}{4 \mathrm{i}} \sqrt{\kappa_{A} \kappa_{B}} \left[ \frac{\partial  }{\partial \beta_{1}} (S_{+} -  S_{-}) + \frac{\partial  }{\partial \beta_{2}} (S_{+} -  S_{-}) \right] ,  \label{eq:derivative 3}\\ 
    \frac{\partial h_{2}}{\partial \beta_{y}}  &=  \frac{1}{4 \mathrm{i}} \sqrt{\kappa_{A} \kappa_{B}}  \left[ \frac{\partial  }{\partial \beta_{1}} (S_{+} -  S_{-}) - \frac{\partial  }{\partial \beta_{2}} (S_{+} -  S_{-}) \right] .  \end{align} 
\end{subequations}
The symmetries of the lattice sums: inversion symmetry $S_{\pm}(\beta_{1},\beta_{2}) = S_{\mp}( - \beta_{1}, -\beta_{2})$ and reflection symmetry $S_{\pm}(\beta_{1},\beta_{2}) = S_{\pm}(\beta_{2},\beta_{1})$, impose corresponding symmetries on the partial derivatives, which are summarized in Table~\ref{table:symmetry of derivatives}.

\begin{table}[t]
\begin{tabular}{lcccc}
\hline
\multicolumn{1}{|l|}{}                    & \multicolumn{1}{c|}{ $\frac{\partial h_{1}}{\partial \beta_{x}}  $ } & \multicolumn{1}{c|}{$\frac{\partial h_{1}}{\partial \beta_{y}}$} & \multicolumn{1}{c|}{$\frac{\partial h_{2}}{\partial \beta_{x}} $} & \multicolumn{1}{c|}{$\frac{\partial h_{2}}{\partial \beta_{y}} $} \\ \hline
\multicolumn{1}{|c|}{   Inversion $\bm{\beta} \to - \bm{\beta}$ }  & \multicolumn{1}{c|}{$-$} & \multicolumn{1}{c|}{$-$} & \multicolumn{1}{c|}{$+$} & \multicolumn{1}{c|}{$+$} \\ \hline
\multicolumn{1}{|c|}{Reflection $\beta_{1} \leftrightarrow \beta_{2}$ } & \multicolumn{1}{c|}{$+$} & \multicolumn{1}{c|}{$-$} & \multicolumn{1}{c|}{$+$} & \multicolumn{1}{c|}{$-$} \\ \hline
                                          & \multicolumn{1}{l}{}   & \multicolumn{1}{l}{}   & \multicolumn{1}{l}{}   & \multicolumn{1}{l}{}  
\end{tabular}
\caption{Symmetry properties of the Hamiltonian derivatives.}
\label{table:symmetry of derivatives}
\end{table}

The Dirac points $K$ and $K^{\prime}$ are fixed points under the combined operation of inversion followed by reflection, as seen in Fig.~\ref{fig:reciprocal lattice}. From Table~\ref{table:symmetry of derivatives}, the derivatives $\partial h_{1}/\partial \beta_{x}$ and $\partial h_{2}/\partial \beta_{y}$ are odd under this combined symmetry and therefore must vanish at these points. In contrast, $\partial h_{1}/\partial \beta_{y}$ and $\partial h_{2}/\partial \beta_{x}$ are even and are generally non-vanishing.

The quasi-inversion symmetry shown in Eq.~\eqref{eqs:quasi-inversion symmetry 3} dictates a crucial relationship between the two non-vanishing derivatives, namely that they are equal and opposite: $-\partial h_{1}/\partial \beta_{y} = \partial h_{2}/\partial \beta_{x}$ at the Dirac points. Using the definitions from Eqs.~\eqref{eq:derivative 2} and \eqref{eq:derivative 3}, we see that this is equivalent to demonstrating the following identity for the lattice sums:
\begin{equation} 
    \left[ {e}^{-\mathrm{i} \frac{2}{3} \pi  } \frac{\partial }{\partial \beta_{1}} - {e}^{\mathrm{i} \frac{2}{3} \pi  } \frac{\partial }{\partial \beta_{2}} \right] S_{+}(\beta_{1},\beta_{2}) \bigg\rvert_{\bm{\beta} = \bm{\beta}_{K}} = \left[ - {e}^{\mathrm{i} \frac{2}{3} \pi  } \frac{\partial }{\partial \beta_{1}} + {e}^{-\mathrm{i} \frac{2}{3} \pi  } \frac{\partial }{\partial \beta_{2}} \right] S_{-}(\beta_{1},\beta_{2}) \bigg\rvert_{\bm{\beta} = \bm{\beta}_{K}}. 
\end{equation}
This relation can be verified by differentiating the quasi-inversion symmetries given in Eq.~\eqref{eqs:quasi-inversion symmetry 3} and using the fact that $S_{+}(\bm{\beta}_{K}) = S_{-}(\bm{\beta}_{K}) = 0$.
Substituting these results into the first-order expansion yields the low-energy effective Hamiltonians. Around the $K$ point, we have:
\begin{equation} H(\bm{\beta}_{K} +\mathbf{q}) = -v_{F} q_{y} \sigma_{1}+ v_{F} q_{x} \sigma_{2}+ \mathcal{O}(\mathbf{q}^{2}). \end{equation}
Similarly, for the $K^{\prime}$ point, we find:
\begin{equation} H(\bm{\beta}_{K^{\prime}} +\mathbf{q}) = v_{F} q_{y} \sigma_{1}+ v_{F} q_{x} \sigma_{2}+ \mathcal{O}(\mathbf{q}^{2}). \end{equation}
These expressions take the canonical form of the 2D massless Dirac equation. This result demonstrates that, at low energies, the system with long-ranged interactions is indistinguishable from the well-known tight-binding model of graphene.

\bibliographystyle{apsrev4-2}
\bibliography{reference}

\end{document}